\documentclass[journal, final]{IEEEtran}

\usepackage[T1]{fontenc}
\usepackage[utf8]{inputenc}
\usepackage{cite}
\usepackage{amsmath, mathtools}
\usepackage{amsfonts,amsthm,bm}
\usepackage{amssymb}
\usepackage{comment}
\usepackage{graphicx}
\usepackage{standalone}
\usepackage{dsfont}
\usepackage{mathtools}
\usepackage{color, colortbl}
\usepackage{soul}
\usepackage[normalem]{ulem}
\usepackage[acronym,shortcuts]{glossaries}
\usepackage{arydshln}
\usepackage{bbm}
\usepackage{svg} 
\usepackage{algorithm,algorithmic}
\usepackage{adjustbox}
\usepackage[inline]{enumitem}
\usepackage{multirow}
\usepackage{tabularx}
\usepackage{booktabs}

\usepackage{comment}
\usepackage{tikz}
\usepackage{pgfplots}
\pgfplotsset{compat=newest}
\pgfplotsset{plot coordinates/math parser=false}
\newlength\fheight
\newlength\fwidth
\usetikzlibrary{plotmarks,patterns,decorations.pathreplacing,backgrounds,calc,arrows,arrows.meta,spy,matrix}
\usepgfplotslibrary{patchplots,groupplots}
\usepackage{tikzscale}

\usetikzlibrary {patterns.meta}

\DeclareMathOperator*{\argmax}{\arg\!\max}
\DeclareMathOperator*{\argmin}{\arg\!\min}

\DeclareMathOperator*{\diag}{diag}

\newtheorem{theorem}{Theorem}

\usepackage{ulem}

\providecolor{added}{rgb}{0,0,1}
\providecolor{deleted}{rgb}{1,0,0}
\newacronym{3gpp}{3GPP}{3rd Generation Partnership Project}
\newacronym{5g}{5G}{fifth-generation}
\newacronym{6g}{6G}{sixth-generation}
\newacronym{af}{AF}{amplify-and-forward}
\newacronym{cdf}{CDF}{cumulative distribution function}
\newacronym{csi}{CSI}{channel state information}
\newacronym{cwc}{CWC}{capacity-weighted clustering}
\newacronym{dlos}{dLoS}{deterministic \ac{los}}
\newacronym{ecdf}{ECDF}{empirical cumulative distribution function}
\newacronym{fdma}{FDMA}{frequency division multiple access}
\newacronym{fpga}{FPGA}{field programmable gate array}
\newacronym{fov}{FoV}{field-of-view}
\newacronym{gnb}{gNB}{next generation Node Base}
\newacronym{hc}{HC}{hierarchical clustering}
\newacronym{km}{KM}{K-means}
\newacronym{kmed}{KMed}{K-medoids}
\newacronym{iab}{IAB}{integrated access and backhaul}
\newacronym{icwc}{ICWC}{inverse capacity-weighted clustering}
\newacronym{irs}{IRS}{intelligent reflecting surface}
\newacronym{isd}{ISD}{inter-site distance}
\newacronym{los}{LoS}{line-of-sight}
\newacronym{lsfc}{LSFC}{large-scale fading coefficient}
\newacronym{minlp}{MINLP}{mixed integer nonlinear programming}
\newacronym{mimo}{MIMO}{multiple input multiple output}
\newacronym{mmwave}{mmWave}{millimeter wave}
\newacronym{nlos}{NLoS}{non-line-of-sight}
\newacronym{noma}{NOMA}{non-orthogonal multiple access}
\newacronym{nr}{NR}{new radio}
\newacronym{oscbc}{OSCBC}{one-shot capacity-based clustering}
\newacronym{ofdma}{OFDMA}{orthogonal frequency-division multiple access}
\newacronym{oma}{OMA}{orthogonal multiple access}
\newacronym{pam}{PAM}{partition around medoids}
\newacronym{plos}{pLoS}{probabilistic \ac{los}}
\newacronym{qos}{QoS}{Quality of Service}
\newacronym{rb}{RB}{resource block}
\newacronym{rsma}{RSMA}{rate-splitting multiple access}
\newacronym{scm}{SCM}{spatial channel model}
\newacronym{snr}{SNR}{signal-to-noise-ratio}
\newacronym{sinr}{SINR}{signal-to-interference-plus-noise-ratio}
\newacronym{siso}{SISO}{single input single output}
\newacronym{svd}{SVD}{singular value decomposition}
\newacronym{tdma}{TDMA}{time division multiple access}
\newacronym{tti}{TTI}{transmission time interval}
\newacronym{thz}{THz}{Terahertz}
\newacronym{ue}{UE}{user equipment}
\newacronym{ula}{ULA}{uniform linear array}
\newacronym{uma}{UMa}{urban macro-cell}
\newacronym{umi}{UMi}{urban micro-cell}
\newacronym{upa}{UPA}{uniform planar array}

\title{Downlink Clustering-Based Scheduling of IRS-Assisted Communications With Reconfiguration Constraints}

\author{Alberto Rech, Matteo Pagin, Leonardo Badia, Stefano Tomasin,\\ Marco Giordani, Jonathan Gambini, Michele Zorzi 
\thanks{Alberto Rech, Leonardo Badia, Matteo Pagin, Stefano Tomasin, Marco Giordani, and Michele Zorzi are with the Department of Information Engineering, University of Padova, Italy.
Jonathan Gambini is with the Milan Research Center, HUAWEI, Italy. A. Rech is also with the Paris Research Center, HUAWEI, France.
Corresponding author: A. Rech (email: alberto.rech@huawei.com).}
\thanks{%This work was supported in part by the European Union through the Italian National Recovery and Resilience Plan (NRRP) of NextGenerationEU, partnership on ``Telecommunications of the Future'' (Program ``RESTART'') under Grant PE0000001. 
Project funded under the National Recovery and Resilience Plan (NRRP), Mission 4 Component 2 Investment 1.3 - Call for tender No. 341 of 15 marzo 2022 of Italian Ministry of University and Research funded by the European Union – NextGenerationEU Project code MUR: PE\_0000001 Concession Decree No. 1549 of 11/10/2022 adopted by the Italian Ministry of University and Research, CUP C93C22005250001 Project title RESearch and innovation on future Telecommunications systems and networks, to make Italy more smART (RESTART).} 
\thanks{Part of this paper has been presented at the IEEE Wireless Communications and Networking Conference (WCNC), [1]].}}

\begin{document}
\maketitle

\bstctlcite{IEEEexample:BSTcontrol}
\nocite{rech2023downlink}

\begin{abstract}
\Acp{irs} are being widely investigated as a potential low-cost and energy-efficient alternative to active relays for improving coverage in next-generation cellular networks. However, technical constraints in the configuration of \acp{irs} should be taken into account in the design of scheduling solutions and the assessment of their performance.
To this end, we examine an \ac{irs}-assisted \ac{tdma} cellular network where the reconfiguration of the \ac{irs} incurs a communication cost; thus, we aim at limiting the number of reconfigurations over time. 
Along these lines, we propose a clustering-based heuristic scheduling scheme that maximizes the cell sum capacity, subject to a fixed number of reconfigurations within a \ac{tdma} frame. 
First, the best configuration of each \ac{ue}, in terms of joint beamforming and optimal \ac{irs} configuration, is determined using an iterative algorithm. Then, we propose different clustering techniques to divide the \acp{ue} into subsets sharing the same sub-optimal \ac{irs} configuration, derived through distance- and capacity-based algorithms. Finally, \acp{ue} within the same cluster are scheduled accordingly.  
We provide extensive numerical results for different propagation scenarios, \ac{irs} sizes, and phase shifters quantization constraints, showing the effectiveness of our approach in supporting multi-user \ac{irs} systems with practical constraints.
\end{abstract}


\glsresetall
\IEEEpeerreviewmaketitle

\section{Introduction}
\label{sec:introduction}

The ever-increasing growth of mobile traffic has called both academia and industry to identify and develop solutions for extending the radio spectrum beyond the crowded sub-6 GHz bands. As a result, the use of \ac{mmwave} band for cellular communications has been included in the latest releases of the \ac{3gpp}  standard, namely \ac{5g} New Radio (NR)~\cite{3gpp.38.104}. Moreover, \ac{thz} frequencies are being investigated as enablers for \ac{6g} networks as well~\cite{tariq2020speculative}.

However, transmissions in the \ac{mmwave} and \ac{thz} bands are subject to challenging propagation conditions, mainly due to severe path loss and susceptibility to blockages~\cite{rangan2017potentials}.
To mitigate these limitations, the research community has explored solutions to improve network coverage, for example using \ac{iab} nodes, as also approved by the \ac{3gpp} as part of \ac{5g} \ac{nr} specifications for Rel-16~\cite{3gpp.38.174}.
In particular, \ac{iab} allows base stations, or \acp{gnb} in 5G NR parlance, to establish wireless (rather than traditional fiber-like) backhaul links, possibly through multiple hops,  to a donor, thus reducing deployment costs~\cite{polese2020integrated}. Still, \ac{iab} involves complex signal processing and saturation of the available resources and may be costly and energy-consuming for network operators.

In light of this, \acp{irs} are being investigated as solutions to overcome the harsh propagation conditions shown by \ac{mmwave} and \ac{thz} bands in a cost- and energy-efficient manner~\cite{flamini2022towards}. 
\Acp{irs} are meta-surfaces, whose radiating elements can \emph{passively} tune the phase shift of impinging signals to favorably alter an electromagnetic field towards an intended destination. They can be configured to beamform the reflected signal virtually in any direction, hence acting as a relay to improve the signal quality without an active (power-consuming) amplification~\cite{bjornson2019intelligent}. 

\subsection{Prior Work}
Despite the substantial research hype, most recent studies on \acp{irs} rely on strong assumptions that do not match real-world deployments.
Specifically, a significant body of literature is based on the assumption that \acp{irs} establish an ideal (i.e., fiber-like) control channel with the base station~\cite{wu2020towards, abeywickrama2020intelligent, wu2019intelligent, pagin2022end}. Instead, actual deployments are expected to feature a wireless, i.e., error-prone, \ac{irs} control, possibly implemented with low-cost technologies~\cite{liu2022path, liaskos2018realizing}. 
This introduces constraints on the \ac{irs} reconfiguration period, which needs to be synchronized with the base station to beamform the signal towards the \ac{ue} served during the specific time slot~\cite{flamini2022towards}, a similar research problem to scheduling in cellular networks. 

In this perspective, \ac{irs}-assisted downlink scheduling solutions have been widely studied in different domains, each with its own theoretical constraints.
For example, in \ac{ofdma} user scheduling, all the users scheduled in a given time slot must be served using the same reflection coefficients, due to the lack of frequency selective beamforming capabilities at the \ac{irs}. In this context, dynamic optimization schemes, wherein the \ac{irs} configurations are adjusted at each time slot, have been studied in ~\cite{Yang20IRS, Lee23Harmony}.
The authors of \cite{guo2021intelligent} consider a two-user downlink transmission problem in an \ac{irs}-assisted scenario over fading channels, and compare the results of different basic \ac{oma} and \ac{noma} schemes. It is found that, while \ac{noma} is the best solution, by exploiting \ac{irs} reconfiguration in each slot of the fading block, \ac{tdma} outperforms \ac{fdma}, and its performance is similar to that of \ac{noma}.
A hybrid \ac{tdma}-\ac{noma} approach, instead, was investigated in
an uplink scenario in \cite{zhang2021throughput, alobiedollah2023self}, in the context of a wireless-powered network, where users are grouped based on their channel gains. 
Then, \acp{ue} within the same group transmit in a non-orthogonal fashion, while different groups are assigned to different time slots.
Moreover, a user scheduling algorithm based on graph neural networks, able to jointly optimize the \ac{irs} configuration and the \ac{gnb} beamforming in downlink, was recently presented in~\cite{Zhang22Learning}. 
Similarly, the authors of~\cite{Bansal21Rate, Fu21Resource, Zhuo22Partial} evaluated the performance of several non-orthogonal downlink scheduling methods, such as \ac{rsma}.
Finally, \acp{irs} with energy harvesting capabilities are considered in~\cite{hu2021robust}. In this work, the authors propose a trade-off between the system sum capacity and the \ac{irs} energetic self-sustainability, with the goal of achieving coverage flexibility and low deployment costs.

Still, most of the literature poses little to no reconfiguration constraints for the IRS.
However, early \ac{irs} control circuitry prototypes, which have low power consumption (i.e., a few hundreds of mW), have a non-negligible phase-shifts reconfiguration time \cite{mu2021capacity, ETSIGRRIS003}, thus posing additional constraints in the system design. For example, the prototypes in \cite{rossanese2022designing} and \cite{alexandropoulos2023ris} have a reconfiguration time of a few tens of ms, even though architectures based on \ac{fpga} such as in \cite{yezhen2020novel} promise to achieve much lower configuration times, i.e., in the order of tens of microseconds.
Still, the overhead (in terms of time)  increases as the number of IRS elements increases, as investigated in \cite{jamali2022low, nadeem2020asymptotic, Qian22joint}. 
In any case, a constraint on the number of reconfigurations (and relative period) is desirable to ensure system synchronization and minimize the IRS downtime during reconfiguration. In this regard, it is of interest to 
\begin{enumerate*}[label=(\textit{\roman*})]
  \item investigate the level of performance degradation experienced by \ac{irs}-assisted systems when considering practical constraints, including limitations in the number of reconfigurations, and
  \item design algorithms that can mitigate these constraints. 
\end{enumerate*}
The limitation on the number of \ac{irs} reconfigurations in a given time frame has been initially studied in \cite{mu2021capacity}, where the authors evaluate the capacity of both \ac{oma} and \ac{noma} schemes of a 2-user \ac{irs}-assisted \ac{siso} system under Rayleigh fading conditions. Still, additional research efforts is required to fully characterize the impact of IRS reconfigurations constraints on the network.

\subsection{Contributions} 
In this paper, we propose a \ac{tdma} scheduling policy for downlink cellular transmissions based on clustering algorithms, to maximize the sum capacity in \ac{irs}-assisted network deployments with practical constraints.
Notably, we assume a fixed maximum number of \ac{irs} reconfigurations within a time frame and aim at optimizing both the reconfiguration time and the resulting \ac{irs} configuration(s).
The limit on the number of reconfigurations sets a simple constraint on the overhead entailed by the control of the \ac{irs}. 
Our contributions can be summarized as follows.
First, we formalize an optimization problem to determine the optimal \ac{irs} configuration(s) to maximize the sum capacity. Specifically, we partition the \acp{ue} into disjoint groups, and propose an algorithm that iteratively optimizes the \ac{irs} configuration for each group, as well as the relative beamformers to be used for transmission.
Secondly, we convert the sum capacity problem into a clustering problem to determine the optimal set of groups, or clusters, for the \acp{ue} based on their channel characteristics. 
Accordingly, we design clustering algorithms for the \acp{ue}, such that all \acp{ue} in the same cluster will adopt the same \ac{irs} configuration. The goal is to minimize the capacity loss associated with a sub-optimal \ac{irs} configuration for each \ac{ue} in the cluster, for promoting communication efficiency and reducing the overhead.
We investigate two clustering techniques: 
\begin{enumerate*}[label=(\textit{\roman*})]
\item distance-based clustering, which adjusts the cluster configuration based on the distance with the optimal \ac{irs} configuration of each \ac{ue} in the cluster;
\item capacity-based clustering, which adjusts the cluster configuration to maximize the sum capacity and/or the user fairness.
 Specifically, we propose three clustering algorithms. The first, named \ac{cwc}, iteratively adjusts the clusters' centroids weighting the points in each cluster by the capacity achieved by each \ac{ue} in the cluster, therefore favoring the best users and maximizing the sum capacity. The second, named \ac{oscbc}, is a low-complexity approach where the centroids are simply the optimal configurations of the \acp{ue} experiencing the highest \ac{snr} in each cluster, without considering the impact of the remaining \acp{ue}.  Finally, \ac{icwc} promotes fairness among the \acp{ue} by weighting the points in the clusters by the inverse of the \acp{ue}' achievable rate.
 \end{enumerate*}

Moreover, we compare via simulation the performance of the distance- and capacity-based clustering in different \ac{irs}-assisted scenarios.
Extensive numerical results show that there exists a trade-off between the achievable sum capacity and fairness, even though capacity-based clustering techniques can guarantee satisfactory performance, despite some reconfiguration constraints. Also, we demonstrate that scheduling based on clustering can reduce by 50\% the number of \ac{irs} reconfigurations, thus promoting communication efficiency, in view of minor degradation in terms of sum capacity. 

Finally, with respect to \cite{rech2023downlink}, we introduce new capacity-based clustering strategies to improve fairness and provide more extensive numerical results to demonstrate the scalability of the proposed solutions as a function of the density of \acp{ue} and the \ac{irs} size. Moreover, we evaluate the performance of the proposed scheduling strategies considering realistic \ac{irs} network constraints, including the quantization of phase shifts, and for different channel propagation conditions.
In this sense, we provide additional results in terms of the computational complexity of the proposed distance- and capacity-based clustering algorithms, as well as in terms of fairness.

\subsection{Organization and Notation}
The rest of the paper is organized as follows. In Section \ref{sec:system_model}, we introduce the system model. In Section~\ref{sec:sumcapoptimization}, we present the sum capacity optimization problem. In Section \ref{sec:optimization}, we describe the scheduling framework, while in Sections \ref{sec:dist_based} and \ref{sec:cap_based} we present distance-based and capacity-based clustering algorithms, respectively. In Section~\ref{sec:numerical_results}, we show numerical results and compare the different scheduling and clustering solutions. Finally, Section~\ref{sec:conclusions} draws the main conclusions.

Scalars are denoted by italic letters; vectors and matrices by boldface lowercase and uppercase letters, respectively; sets are denoted by calligraphic uppercase letters. $\diag(\bm{a})$ indicates a square diagonal matrix with the elements of $\bm{a}$ on the principal diagonal. $\bm{A}^{\rm T}$ and $\bm{A}^\dagger$ denote the transpose and the conjugate transpose of matrix $\bm{A}$, respectively. $[\bm{A}]_{k \ell}$ denotes the scalar value in the $k$-th row and $\ell$-th column of matrix $\bm{A}$, while $[\bm{a}]_k$ denotes the $k$-th element of vector $\bm{a}$. The imaginary unit is denoted as $j =\sqrt{-1}$, and $\angle a$ denotes the phase of $a \in \mathbb{C}$. Finally, $\mathbb{E}[\cdot]$ denotes statistical expectation.

\section{System Model}
\label{sec:system_model}

\begin{figure}[t]
    \centering
    \includegraphics[width=\linewidth, trim={2.8cm 0 1.8cm 0},clip]{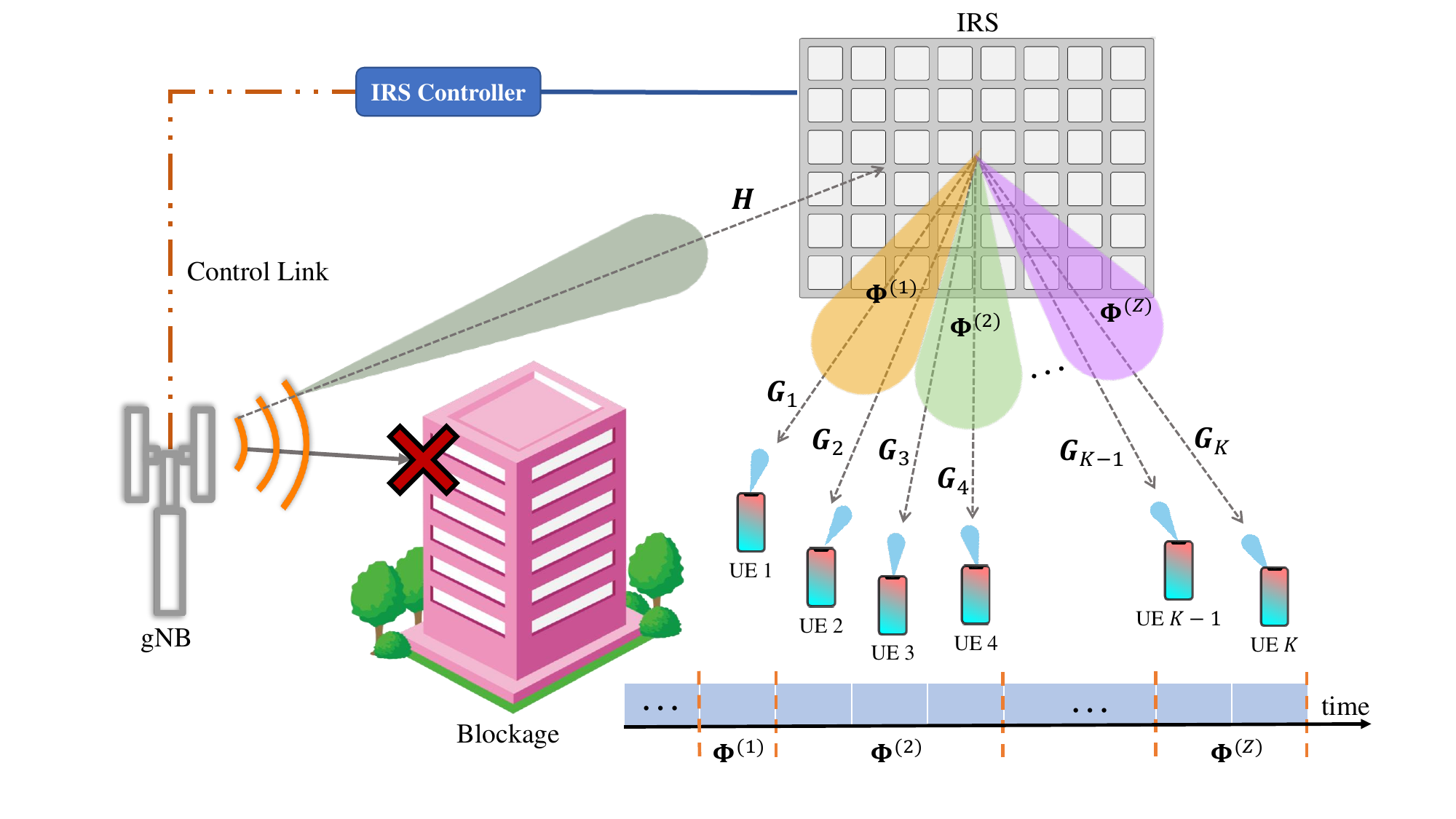}    
    \caption{Downlink \ac{tdma} scheduling for multi-user \ac{irs}-aided systems.}
    \label{fig:system_model}
\end{figure}

We consider downlink data transmissions for the multi-user \ac{mimo} communication system shown in Fig.~\ref{fig:system_model}, wherein the transmission from the \ac{gnb} to the $K$ \acp{ue} is assisted by an \ac{irs}.
The \ac{gnb} and the \acp{ue} are equipped with $N_{\rm g}$ and $N_{\rm U}$ antennas, respectively. We assume that the direct link between the \ac{gnb} and the \acp{ue} is unavailable due to blockage. As a consequence, the \ac{gnb} transmits signals to the \acp{ue} by exploiting the virtual link offered by the \ac{irs}. In this context, the \ac{irs} configuration is managed by the \ac{gnb} through the \ac{irs} controller, by exploiting a dedicated link between the \ac{gnb} and the \ac{irs}, thus with no additional communication overhead in the \ac{gnb}-\ac{ue} link.
Time is divided into frames of $K$ slots, and each \ac{ue} is served exactly once in a frame in a \ac{tdma} fashion, which ensures there is no co-channel interference as \acp{ue} are separated in the time domain.
We assume that \acp{ue} are either static or moving slowly, which is the most typical application scenario for \ac{irs}-assisted networks. Under such conditions, the channel coherence time is in the order of 10~ms~\cite[Fig.~5]{lu2020positioning}. Considering that perfect \ac{csi} of all \acp{ue} is acquired at the \ac{gnb} at the beginning of each frame (a realistic assumption that does not affect the proposed scheduling framework for IRS communication), it is reasonable to conclude that the channel remains constant throughout the whole time frame. Here, we assume that the CSI is available for any IRS configuration.

\subsection{\ac{irs} Model}\label{sec:irsmodel}
Each of the $N_{\rm I}$ elements of the \ac{irs} acts independently as an omnidirectional antenna unit that reflects the impinging electromagnetic field by introducing a tunable phase shift on the baseband-equivalent signal. We denote as $\phi_n=e^{j\theta_n}$ the reflection coefficient of the $n$-th \ac{irs} element, where $\theta_n \in \mathcal{P}_{\theta}$ is the induced phase shift, and $\mathcal{P}_{\theta}$ is the set of possible phase shifts. 
Recent works argue that continuous phase shifts are hardly implementable in practice \cite{Tan2018}. Therefore, we consider both continuous and quantized phase shifts. While in the former case the set of phase shifts is $\mathcal{P}_{\theta}= [-\pi, \pi)$, in the latter we have 
$\mathcal{P}_{\theta} = \left\{0, \frac{2\pi}{2^b},\ldots,\frac{2\pi(2^b-1)}{2^b} \right\}$ where $b>0$ is the number of bits employed to quantize the phase shifts. 

We denote with $\bm{H} \in \mathbb{C}^{N_{\rm I} \times N_{\rm g}}$ the channel matrix between the \ac{irs} and the gNB, and with $\bm{G}_k \in \mathbb{C}^{N_{\rm U} \times N_{\rm I}}$ the channel matrix of the link between the \ac{irs} and \ac{ue} $k$, respectively.
We consider single-stream transmissions,\footnote{The assumption of single-stream transmissions is justified by the rank of the cascade channel matrix, which is likely equal to one. This conclusion comes from the considerations reported in \cite{he2020cascaded, rains2023ris,rappaport2017investigation}, and has been verified numerically for the considered setup.} with $\bm{w}_k \in \mathbb{C}^{N_{\rm g}\times 1}$ and $\bm{v}_k\in \mathbb{C}^{N_{\rm U}\times 1}$ defined as the beamforming vectors at the \ac{gnb} and \ac{ue} $k$, respectively.
Let $x_k$ be the single-stream signal transmitted by the \ac{gnb} to \ac{ue} $k$; the received post-processing signal can be expressed as
\begin{equation}
    z_k = \bm{v}_k^{\rm T} \bm{G}_k \bm{\Phi} \bm{H}  \bm{w}_kx_k + \bm{v}_k^{\rm T} \bm{n}_{k},
\end{equation}
where $\bm{n}_k\in \mathbb{C}^{N_{\rm U}\times 1}$ represents the circularly symmetric complex Gaussian noise vector with entries having zero mean and variance $\sigma^2_{n}$, while $\bm{\Phi} \in \mathbb{C}^{N_{\rm I}\times N_{\rm I}}$ is the {\ac{irs} configuration}, i.e., a diagonal matrix defined as $\bm{\Phi} = \diag(\phi_1,\ldots,\phi_{N_{\rm I}})$.
Note that different, and specific, \ac{irs} configurations can be adopted for different \acp{ue}. Accordingly, in the rest of the paper we let $\bm{\Phi}_k$ be the \ac{irs} configuration adopted when \ac{ue} $k$ is served.

The \ac{snr} at \ac{ue} $k$ under \ac{irs} configuration $\bm{\Phi}_k$ is
\begin{equation}\label{snr_k}
    \Gamma_k(\bm{\Phi}_k) = \frac{|\bm{v}_k^{\rm T} \bm{G}_k \bm{\Phi}_k \bm{H}  \bm{w}_k |^2\sigma_{x}^2}{|\bm{v}_k|^2\sigma_{n}^2},
\end{equation}
where $\sigma_{x}^2$ is the power of the transmitted signal.
To maximize the \ac{snr} of a given \ac{ue}, a specific \ac{irs} configuration should be adopted, tailored to the \ac{ue} position in the cell and the channel conditions.
However, the goal of this paper is to limit the number of \ac{irs} reconfigurations to comply with realistic overhead constraints, as well as to improve the communication efficiency, and algorithms seeking to comply with these requirements will be presented in Section~\ref{sec:optimization}.

\section{Sum Capacity Optimization Problem}\label{sec:sumcapoptimization}

We impose a constraint on the number of \ac{irs} reconfigurations per time frame, with the goal of either limiting the reconfiguration,\footnote{We remark that the \ac{gnb} typically communicates a (possibly new) \ac{irs} configuration in each \ac{tti}. The reconfiguration constraint introduced in the proposed IRS scheduling framework is able to reduce this overhead by a factor ${Z}/{K} \leq 1$.} for accounting for practical limitations that might arise in realistic deployments.
On the downside, achieving this objective usually leads to \ac{snr} degradation as sub-optimal \ac{irs} configurations might be adopted for some \acp{ue}. To mitigate this effect, we formulate a constrained optimization problem on the average cell sum capacity.
Specifically, we assume the following conditions:
\begin{enumerate}[label=\arabic*.]
    \item at most $Z$ \ac{irs} reconfigurations can occur per time frame;
    \item the \ac{gnb} serves $K$ \acp{ue} by partitioning them into $Z$ disjoint subsets $\mathcal{U}_1,\ldots,\mathcal{U}_{Z}$, $Z\leq K$;
    \item for each \ac{ue} in $\mathcal{U}_z$, the same \ac{irs} configuration $\bm{\Phi}^{(z)}$ is used, i.e., $\bm{\Phi}_k = \bm{\Phi}^{(z)}, \forall \, k \in \mathcal{U}_z, \forall \, 1\leq z \leq Z$.
\end{enumerate}

Then, the achievable rate of \ac{ue} $k\in \mathcal{U}_z$ is
\begin{equation}
    R_k(\bm{\Phi}^{(z)}) = \log_2\left(1+\Gamma_k\big(\bm{\Phi}^{(z)}\big)\right),
    \label{eq:rate}
\end{equation}
where $\Gamma_k(\bm{\Phi}^{(z)})$ is the \ac{snr} experienced by the $k$-th \ac{ue} while configuration $\bm{\Phi}^{(z)}$ is adopted at the \ac{irs}, i.e., the configuration shared by all \acp{ue} belonging to subset $\mathcal{U}_z$.

Let $\mathcal{I} = \{\bm{\Phi}^{(1)}, \bm{\Phi}^{(2)}, \ldots, \bm{\Phi}^{(Z)}\}$ be the set of \ac{irs} configurations corresponding to subsets $\mathcal{U}_1,\ldots,\mathcal{U}_{Z}$. The system sum capacity within a time frame is defined as
\begin{equation}\label{eq:sumrate}
    C(\mathcal{U}_1,\ldots,\mathcal{U}_{Z},\mathcal{I}) = B\sum_{z=1}^{Z} \sum_{k \in \mathcal{U}_z}R_k\big(\bm{\Phi}^{(z)}\big),
\end{equation} 
where $B$ is the transmission bandwidth. The optimization problem is then formulated as
\begin{subequations}\label{optproblem}
    \begin{equation}
        \max_{\substack{\mathcal{U}_1,\ldots,\mathcal{U}_{Z}, \mathcal{I}}} \quad  C(\mathcal{U}_1,\ldots,\mathcal{U}_{Z},\mathcal{I}),
        \\    
    \end{equation}
    \begin{alignat}{2}
     & \text{s.t.}\; 
     \quad & \angle\big[\bm{\Phi}^{(z)}\big]_{n,n} \in \mathcal{P}_{\theta},\quad \forall\, n, z.\label{const_phaseshifters}
    \end{alignat}
\end{subequations}

Problem \eqref{optproblem} determines the optimal grouping strategy for the \acp{ue} subsets $\mathcal{U}_1,\ldots,\mathcal{U}_{Z}$, and assigns the best \ac{irs} configuration accordingly. Therefore, \eqref{optproblem} is both continuous (i.e., the optimization of the IRS configuration) and combinatorial (i.e., the grouping of the \acp{ue}), and can be thus classified as a \ac{minlp} problem.
Moreover, the following theorem holds.
\begin{theorem}
    The sum capacity maximization problem \eqref{optproblem} is NP-complete.
\end{theorem}
\begin{IEEEproof}
    First, we observe that the problem falls within the general NP class. This is because if \eqref{optproblem} is solved to find $\mathcal{U}_1, \ldots, \mathcal{U}_Z, \mathcal{I}$, both the sum capacity and the phase-shift constraints \eqref{const_phaseshifters} could be verified in polynomial time.
    To prove that the problem in NP-complete, we set $\mathcal{I}$, and consider the simplified problem
    \begin{equation}
        \max_{\substack{\mathcal{U}_1,\ldots,\mathcal{U}_{Z}}} \quad  C(\mathcal{U}_1,\ldots,\mathcal{U}_{Z},\mathcal{I}).
    \end{equation}
     This problem can be viewed as a multi-knapsack problem with different clusters $\mathcal{U}_1, \ldots, \mathcal{U}_Z$ as knapsacks, and the goal is to maximize the total system capacity.
     This is known to be NP-hard, as it is a generalization of the classic knapsack problem. 
     The original sum capacity maximization problem \eqref{optproblem}, where we consider the additional degrees of freedom of the \ac{irs} configurations, remains NP-hard, thus making the problem NP-complete.
\end{IEEEproof}

Given the inherent problem complexity, we adopt heuristic clustering algorithms to obtain approximated, though close-to-optimal, solutions, as described in Section~\ref{sec:optimization}.

\section{Heuristic Sum Capacity Maximization}
\label{sec:optimization}
In this section, we provide heuristic solutions to \eqref{optproblem}. First, we present two clustering-based approaches to identify and group \acp{ue} with a similar optimal \ac{irs} configuration. Then, we solve the scheduling problem on the identified clusters with a \ac{tdma} approach \cite{anchora2012capacity}.
 We compute the \acp{ue} clusters by first estimating the optimal {individual \ac{irs} configurations}, denoted as $\bm{\Phi}^*_k$, $1\leq k \leq K$, i.e., the \ac{irs} configurations leading to the maximum capacity for each \ac{ue} $k$, as described in Section~\ref{sec:ind_opt}.
 These configurations would solve \eqref{optproblem} for
$Z = K$, as in this case all \acp{ue} are served in a \ac{tdma} fashion and with their optimal \ac{irs} configuration.
The phase coefficients of the optimal \ac{irs} configuration matrices are then chosen as the initial points of a procedure leveraging clustering algorithms in the $N_{\rm I}$-dimensional space, as explained in Section~\ref{sec:clust_tdma}.

\subsection{Optimal Individual \ac{irs} Configurations  }
\label{sec:ind_opt}
In \ac{mimo} systems, both the \ac{gnb} and the \acp{ue} adopt properly tuned beamformers to match the signal transmissions and receptions to the spatial direction providing the highest channel gain~\cite{flamini2022towards}. For the optimization of the \ac{irs} configuration of each individual \ac{ue}, we adopt a procedure similar to that presented in \cite{Qian22joint}, focusing on single-stream transmissions and, without loss of generality, on \ac{ue} $k$.

For a given \ac{irs} configuration, the optimal beamforming vectors  $\bm{v}_k$ and $\bm{w}_k$ coincide with the singular vectors corresponding to the highest singular value of the wireless channel matrix. In particular, we calculate the \ac{svd} of the overall cascade channel matrix
\begin{equation}\label{eq:svd}
    \bm{G}_k \bm{\Phi}_k \bm{H} = \bm{U}\bm{\Sigma}\bm{V}^\dagger,
\end{equation}
where the right and left singular vectors of $\bm{G}_k \bm{\Phi}_k \bm{H}$ are the columns of $\bm{V}$ and $\bm{U}$, and the corresponding singular values are the diagonal entries of $\bm{\Sigma}$.

In our formulation, the \ac{irs} configuration $\bm{\Phi}_k$ is one of the optimization variables.  Indeed, given $\bm{v}_k$ and $\bm{w}_k$, we can solve
\begin{subequations}
    \begin{equation}\label{optproblem_single}
    \bm{\Phi}^*_k = \argmax_{\substack{\bm{\Phi}_k}} R_k(\bm{\Phi}_k), 
    \end{equation}
    \begin{alignat}{2}
    & \text{s.t.}\; \quad & \angle[\bm{\Phi}_k]_{n,n} \in \mathcal{P}_{\theta},\quad n =1,\ldots,N_{\rm I} \, ,
    \end{alignat}
    \end{subequations}
where $R_k$ is the achievable capacity of \ac{ue} $k$, $1\leq k\leq K$, according to \eqref{eq:rate}.
By defining $\bm{s}_k = \bm{v}_k^{\rm T} \bm{G}$ and $\bm{u}_k = \bm{H} \bm{w}_k$, the signal power can be re-written as
\medmuskip=-1mu
\thickmuskip=-1mu
\begin{equation}
    \left|\bm{s}_k\bm{\Phi}_k \bm{u}_k\right|^2 = \left|\sum_{n = 1}^{N_{\rm I}} |[\bm{s}_k]_{n}||[\bm{u}_k]_{n}|e^{j(\angle[\bm{s}_k]_{n}+\theta_n+\angle[\bm{u}_k]_{n})}  \right|^2.
\end{equation}
\medmuskip=6mu
\thickmuskip=6mu
Then, it is sufficient to observe that the \ac{snr} is maximized when the phase shifts introduced by the \ac{irs} are aligned with the phase shifts accumulated along the various paths, i.e., 
\begin{equation}
\label{eq:theta_opt}
    \theta_{k, n} = -(\angle[\bm{s}_k]_{n} + \angle[\bm{u}_k]_{n}), \quad \forall n.
\end{equation}
Taking into account the possible quantization, the optimal phase shifts are given by
\begin{equation}
    \angle[\bm{\Phi}^*_k]_{n,n} \leftarrow
    \argmin_{\substack{\psi \in \mathcal{P}_{\theta}}}\left(\angle e^{j(\theta_{k, n}-\psi)}\right), \quad \forall n.  
\end{equation}

\begin{algorithm}[H]
\caption{Iterative Alternate \ac{irs} Optimization}\label{alg:ind_opt}
\begin{algorithmic}[1]
\REQUIRE $\bm{G}_k, \bm{H}$
\ENSURE $\bm{\Phi}^*_k$
\STATE $t\gets 0$
\STATE $\bm{v}_k, \bm{w}_k \gets \bm{1}$
\REPEAT
\STATE$\bm{s}_k \gets \bm{v}_k^{\rm T} \bm{G}$, $\bm{u}_k \gets \bm{H}\bm{w}_k$
\STATE$\theta_{k, n} \gets -(\angle[\bm{s}_k]_{n} + \angle[\bm{u}_k]_{n})$
\STATE$\angle[\bm{\Phi}_{k, t}]_{n,n} \gets \argmin_{\substack{\psi \in \mathcal{P}_{\theta}}}(\angle e^{j(\theta_{k, n}-\psi)})$
\STATE$\bm{U}, \bm{\Sigma}, \bm{V}^\dagger \gets$  SVD  of $\bm{v}_k^{\rm T}\bm{G}_k \bm{\Phi}_k \bm{H}\bm{w}_k $
\STATE$\bm{v}_k \gets$ column of $\bm{V}$ corresponding to \\\quad\quad the largest singular value
\STATE$\bm{w}_k \gets$ column of $\bm{U}$ corresponding to \\\quad\quad the largest singular value
\STATE$t \gets t+1$
\UNTIL{$|R_k(\bm{\Phi}_{k, t}) - R_k(\bm{\Phi}_{k, t-1})| < \nu$}
\STATE $\bm{\Phi}^*_k \gets \bm{\Phi}_{k,t}$
\end{algorithmic}
\end{algorithm}

To overcome the interdependence between optimal \ac{irs} configurations and beamforming vectors, we propose an iterative alternate optimization approach. We first estimate the optimal beamforming vectors for a given \ac{irs} configuration using \eqref{eq:svd}. Then, we plug the derived beamformers into \eqref{optproblem_single}, and obtain the corresponding optimal \ac{irs} configuration. We repeat this two-step procedure until convergence, which, for practical purposes, is assumed to be reached when the difference between the achievable rates $R_k$, $\forall k$, in two consecutive iterations is lower than a tolerance $\nu>0$. 
This procedure is summarized in Algorithm~\ref{alg:ind_opt}, where $t$ is the iteration index. The number of iterations grows with the numbers of antennas and \ac{irs} phase shifters. However, from preliminary simulations, and based on the set of parameters we considered (see Section~\ref{sec:numerical_results}), convergence is typically reached in less than $10$ iterations.

\subsection{Clustering-based TDMA Scheduling}
\label{sec:clust_tdma}
For an approximated but close-to-optimal solution to \eqref{optproblem}, we resort to a clustering-based approach.
Our proposed clustering algorithms estimate both the subsets of \acp{ue} $\mathcal{U}_1,\ldots,\mathcal{U}_{Z}$, and the relative set of \ac{irs} configurations $\mathcal{I}$.
We operate on the \textit{phase vector space}, i.e., the points to be clustered are identified by the \ac{irs} phase shifts vector $\left[\angle{\phi_{1}},\ldots,\angle{\phi_{N_{\rm I}}}\right]^{\rm T} = \left[\theta_{1},\ldots,\theta_{N_{\rm I}}\right]^{\rm T}$,
which maps each \ac{irs} configuration $\bm{\Phi}$ to a point in $\left[-\pi,\pi\right)^{N_{\rm I}}$.
In case of quantized phase shifts, the phase vector space is a lattice in the continuous space $\left[-\pi,\pi\right)^{N_{\rm I}}$.

The general clustering-based procedure works as follows:
\begin{itemize}
    \item \emph{Step 1:} find $\bm{\Phi}^*_k,\, \forall k$, i.e., the optimal individual \ac{irs} configurations for each \ac{ue} as in Section \ref{sec:ind_opt};
    \item \emph{Step 2:} build \ac{ue} subsets $\mathcal{U}_z, \, z = 1, \ldots, Z$, by using a clustering algorithm, according to Sections~\ref{sec:dist_based} and \ref{sec:cap_based};
    \item \emph{Step 3:} assign $\bm{\Phi}^{(z)}$ to all \acp{ue} $\in\mathcal{U}_z$.
\end{itemize}

The core idea of this procedure %, and the main contribution of this paper, 
is to use clustering algorithms to group \acp{ue}, and assign the respective \ac{irs} configurations, which are mapped to the \textit{centroid} of the cluster. 
In the case of quantized phase shifts, once the clustering procedure is performed, clusters may share the same centroid and be merged. Therefore, $Z$ represents the \textit{maximum number of clusters}, not the effective number.
Moreover, we remark that the procedure above does not rely on the assumption of perfect \ac{csi}, as the grouping strategy (Step 2) and the individual optimization step (Step 1) are performed independently. Nevertheless, in the case of imperfect \ac{csi}, the estimated individual optimal configurations may differ from the actual optimal configurations, leading to a sub-optimal grouping.

In the following, we propose different techniques to build the clusters based on either a {distance metric} (Section~\ref{sec:dist_based}) or the {achievable rate}  (Section \ref{sec:cap_based}).

\section{Distance-Based Clustering Algorithms}
\label{sec:dist_based}
The class of distance-based clustering contains methods that group data points based on their similarity or dissimilarity according to a distance metric. 
This approach has several advantages, including the efficiency in handling large datasets, and the flexibility to adapt to many different scenarios of interest. 
However, distance-based clustering can be sensitive to the choice of the distance metric (which depends on the nature of the data and the clustering problem), and the initialization values. Moreover, in our specific case, it does not take into account the achievable rate, which is not directly related to the distance among the points in the phase vector space. 

Since the scalar field is the range $[-\pi,\pi)$, the adopted distance has to take into account the circularity of data.
However, the convergence to a local minimum for most of the clustering algorithms is guaranteed only if the points to be clustered belong to a Euclidean space. 
For distance-based algorithms, we thus define the bijective mapping function $f : \mathcal{P}_{\theta}^{N_{\rm I}} \to \mathbb{R}^{2N_{\rm I}}$ as
\begin{align}\label{map_theta}
    f(\bm{\theta}) &= f(\left[\theta_{1},\ldots,\theta_{N_{\rm I}}\right]) \nonumber \\ 
    &= \left[ \cos(\theta_{1}),\sin(\theta_{1}),\ldots,\cos(\theta_{N_{\rm I}}),\sin(\theta_{N_{\rm I}}) \right],
\end{align}
and then define the pairwise {distance} between two generic \ac{irs} configurations $\bm{\alpha}$ and $\bm{\beta}$ as
\begin{equation}\label{circdist}
    \delta(\bm{\alpha},\bm{\beta}) = \vert\vert f(\bm{\alpha}) - f(\bm{\beta}) \vert\vert,
\end{equation}
i.e., the Euclidean distance between the mapping on the unit $N_{\rm I}$-sphere of their respective phase vectors. In the following, with a slight abuse of notation, $f(\bm{\Phi})$ maps the phases of the complex entries in the diagonal of $\bm{\Phi}$ as in \eqref{map_theta}, and $\delta(\bm{\Phi}_1, \bm{\Phi}_2)$ denotes the pairwise distance between the phases of the elements in the diagonal of matrices $\bm{\Phi}_1$ and $\bm{\Phi}_2$.
The sum of squared distances is defined as
\begin{equation}\label{distancemin}
    J(\mathcal{U}_1,\ldots,\mathcal{U}_{Z},\mathcal{I})= \sum_{z=1}^Z\sum_{k \in \mathcal{U}_z}\delta\left(\bm{\Phi}_k^*, \bm{\Phi}^{(z)}\right)^2,
\end{equation}
and the distance-based clustering schemes are used to solve the following problem:
\begin{equation}\label{optproblemdist}
    \min_{\substack{\mathcal{U}_1,\ldots,\mathcal{U}_{Z}, \mathcal{I}}} \quad  J(\mathcal{U}_1,\ldots,\mathcal{U}_{Z},\mathcal{I}),
    \quad \quad \text{s.t.}\quad \eqref{const_phaseshifters}.
\end{equation}

We consider and compare some of the most popular distance-based clustering algorithms, namely, K-means, agglomerative hierarchical clustering, and K-medoids.  

\vspace{2pt}\noindent\emph{\Acl{km}.}
\ac{km} clustering~\cite{rokach2005clustering} aims at finding $Z$ disjoint clusters minimizing the within-cluster squared Euclidean distances. Here, we consider the generalized Lloyd algorithm \cite{Kmeans}, which randomly selects $Z$ points in the space of phase vectors as the initial centroids. In our setup, to ensure optimal performance when $Z=K$, we force the algorithm initialization to a random selection among the phase vectors of the optimal individual \ac{irs} configurations derived in Section~\ref{sec:ind_opt}. Then, in the \textit{assignment step} \ac{km} assigns each data point to the closest centroid, according to the specified distance metric. 
In the subsequent \textit{update step}, the set of centroids is re-computed as the average of the data points that belong to each cluster. These steps are repeated until either convergence or a maximum number of iterations $I_{\rm max}^{\rm KM}$ is reached.
 
\vspace{2pt}\noindent\emph{Agglomerative \acl{hc}.}
The agglomerative \ac{hc}~\cite{murtagh2012algorithms} partitions a set of data points into disjoint clusters by iteratively merging points into clusters until a target number of partitions is met.
In our setup, clusters are initialized as the optimal phase vectors, which thus act as the respective centroids. Then, the average distance between all pairs of data points in any pair of clusters is evaluated. The closest pair of clusters are merged into a new single cluster, whose centroid is computed as the mean of its data points. The procedure is repeated until the number of clusters is $Z$. 

\vspace{2pt}\noindent\emph{\Acl{kmed}.}
\ac{kmed}~\cite{kaufman1987clustering} is a clustering technique similar to \ac{km}, but instead of the mean of the data points within each cluster, it uses the medoid, i.e., the data point that is closest to the center of the cluster. In our setup, we consider the \ac{pam} method~\cite{van2003new}, which starts by randomly selecting $Z$ medoids among the optimal phase vectors and assigns each point to the cluster with the closest medoid. In each iteration, the algorithm evaluates potential \textit{swaps} of medoids with non-medoids. A swap is accepted only if it results in a lower value of the sum of the squared distances to all other data points within the same cluster. The algorithm continues until the medoids no longer change.

\begin{theorem}
    The proposed distance-based clustering techniques converge to a local minimum of \eqref{distancemin}.
\end{theorem}
\begin{IEEEproof}[Proof for KM (LLoyd)]
    In the assignment step, each \ac{ue} $k$ is assigned to the cluster $z$ that minimizes the squared distance $\delta\left(\bm{\Phi}_k^*, \bm{\Phi}^{(z)}\right)^2$. This guarantees that the total sum $J(\mathcal{U}_1,\ldots,\mathcal{U}_{Z},\mathcal{I})$ does not increase. 
    Then, in the update step, $\bm{\Phi}^{(z)}$ is recalculated as the average $\bm{\Phi}_k^*$ within each cluster, so as to minimize the intra-cluster sum of squared distances $\sum_{k \in \mathcal{U}_z}\delta\left(\bm{\Phi}_k^*, \bm{\Phi}^{(z)}\right)^2$, for all $z$.
    Therefore, the conditions of~\cite[Lemma 5]{Sabin1986Global} are satisfied, which ensures  the convergence to a local minimum.\footnote{Notice that~\cite{Sabin1986Global} does not specify the number of iterations needed to reach convergence, which could be large in the case of highly dimensional spaces. Therefore, in practice, we limit the maximum number of iterations to $I_{\rm max}^{\rm KM}$.}
\end{IEEEproof}
\begin{IEEEproof}[Proof for Agglomerative HC]
    At each step, clusters are merged to minimize the increase of the total intra-cluster sum of squared distances. This is equivalent to choosing the merged cluster that results in the smallest increase of $J(\mathcal{U}_1,\ldots,\mathcal{U}_{Z},\mathcal{I})$.
    Then, as in the update step of \ac{km}, the average of the data points minimizes $\sum_{k \in \mathcal{U}_z}\delta\left(\bm{\Phi}_k^*, \bm{\Phi}^{(z)}\right)^2$, for all $z$. Once the number of clusters $Z$ reaches the desired value, convergence to a local minimum is reached.
\end{IEEEproof}
\begin{IEEEproof}[Proof for K-Medoids (PAM)]
    Since, at each iteration, a swap is performed only when it leads to a lower value of the intra-cluster sum of squares, $J(\mathcal{U}_1,\ldots,\mathcal{U}_{Z},\mathcal{I})$ does not increase over different iterations. Given the finite number of data points and possible configurations, the algorithm is guaranteed to converge to a configuration where no swap can further decrease the objective function, thus reaching a local~minimum.
\end{IEEEproof}

\section{Capacity-Based Clustering Algorithms}
\label{sec:cap_based}
The distance-based clustering techniques presented in Section~\ref{sec:dist_based} do not directly take into account the actual capacity achievable by the \acp{ue}, which is a crucial factor for the sum capacity maximization~\eqref{optproblem}. Thus, in the following we propose original capacity-based clustering algorithms that go beyond the state of the art, namely \ac{cwc} (Section~\ref{sec:cwc}), \ac{oscbc} (Section~\ref{sec:clust_oscbc}), and \ac{icwc} (Section~\ref{sec:icwc}).

\subsection{Capacity-Weighted Clustering (CWC)}
\label{sec:cwc}

Similarly to distance-based clustering, also \ac{cwc} proceeds iteratively. However, the stopping condition is based on the variation of the sum capacity of each cluster, rather than on the distance between the centroids. 
In this approach, the clustering algorithm itself weighs the \acp{ue} based on their achievable capacity, so that the parameters of the resulting clusters are closer to those preferred by the \acp{ue} with higher rates, thus promoting the maximization of the sum capacity.

Let $\bm{\Phi}^{(z)}_{i}$ be the \ac{irs} configuration of cluster $\mathcal{U}_{z, i}$ at iteration $i$.
\Acp{ue} are initially sorted in decreasing order of achievable rate.  The algorithm then selects the $Z$ \acp{ue} providing the highest achievable capacity based on the expression in \eqref{eq:rate} with their optimal \ac{irs} configurations.  Without loss of generality, we let $z=1, \dots, Z$ be the  index of those \acp{ue}, and set $\bm{\Phi}_1^{(z)} = \bm{\Phi}^*_z$, $\forall\; 1\leq z\leq Z$, as the centroids of the initial clusters $\mathcal{U}_{1, 0}, \ldots, \mathcal{U}_{Z, 0}$. 
In the following, for simplicity, we denote with $z_{k, i}$ the cluster such that $k \in \mathcal{U}_{z, i}$.  
Each \ac{ue} $k > Z$ is assigned to the cluster whose centroid provides the lowest rate difference with respect to its ideal configuration. Let $R_k(\bm{\Phi}^*_k)$ be the maximum achievable rate of \ac{ue} $k$, obtained from the solution of problem \eqref{alg:ind_opt}. \ac{ue} $k$ is assigned to cluster
\begin{equation}
\label{cluster_assignment}
    z_{k, i} = \argmin_{\substack{z}} [R_k(\bm{\Phi}^*_k) - R_k(\bm{\Phi}^{(z)}_{i})] , 
\end{equation}
where $R_k(\bm{\Phi}^{(z)}_i)$ is the rate achieved by \ac{ue} $k$ adopting the \ac{irs} configuration of cluster $z$ at iteration $i$.
Note that, despite being always non-negative, the rate difference in \eqref{cluster_assignment} cannot be considered a distance metric as, in general, it does not satisfy the triangle inequality.
However, we prove that, as the distance from the optimal configuration increases, the corresponding rate decreases, thus supporting the use of the rate difference as a clustering criterion.

\begin{theorem}
Given the optimal IRS configuration $\bm{\Phi}^*_k$, the rate $R_k(\bm{\Phi})$ is monotonically decreasing with respect to the magnitude of any phase shifts error $\epsilon$.
\end{theorem}
\begin{IEEEproof}
    Let $\epsilon\in [-\pi, \pi]$ be an arbitrary error phase shift, and consider the configuration $\bm{\Phi}_k^\epsilon = \bm{\Phi}^*_k \bm{E}$, where $\bm{E} = \diag(e^{j\epsilon}, 1,\ldots,1)$, i.e., the sub-optimal configuration where only the first \ac{irs} element is affected by the error $\epsilon$.
Assuming, without loss of generality, that $N_{\rm g}=N_k=1$ and $\sigma_x=\sigma_n=1$, the rate $R_k(\bm{\Phi}_k^\epsilon)$ is proportional to $\Gamma_k(\bm{\Phi}_k^\epsilon)$ when using configuration $\bm{\Phi}_k^\epsilon$. The \ac{snr} $\Gamma_k(\bm{\Phi}_k^\epsilon)$ can be written~as
\begin{align}
\label{eq:rate_error}
\Gamma_k(\bm{\Phi}_k^\epsilon) &= |\bm{g}_k \bm{\Phi}^*_k\bm{E} \bm{h}|^2 \\
&= \left|[\bm{g}_k]_1 [\bm{\Phi}^*_k\bm{E}]_{1,1} [\bm{h}]_1 + A \right|^2,
\end{align}
where $A = \sum_{n=2}^{N_{\rm I}}[\bm{g}_k]_n [\bm{\Phi}^*_k]_{n,n} [\bm{h}]_n$. Since $\bm{\Phi}^*_k$ is the optimal configuration, it satisfies \eqref{eq:theta_opt}. It follows that $A \in \mathbb{R}^+$, so \eqref{eq:rate_error} can be further manipulated into
\begin{align}
& \Gamma_k(\bm{\Phi}_k^\epsilon) = | |[\bm{g}_k]_1||[\bm{h}]_1|e^{j \epsilon} + A |^2 \\  
&= A^2 + (|[\bm{g}_k]_1||[\bm{h}]_1|)^2 + 2A|[\bm{g}_k]_1||[\bm{h}]_1|\cos(\epsilon).
\end{align}
Finally, we evaluate the sign of the derivative of $\Gamma_k(\bm{\Phi}_k^\epsilon)$ with respect to the error $\epsilon$ as
\begin{equation}
   \frac{\partial\Gamma_k(\bm{\Phi}_k^\epsilon)}{\partial\epsilon} = -2A|[\bm{g}_k]_1||[\bm{h}]_1|\sin(\epsilon),
\end{equation}
and observe that $\Gamma_k(\bm{\Phi}_k^\epsilon)$, and therefore $R_k(\bm{\Phi}_k^\epsilon)$, is strictly decreasing for $0< |\epsilon|\leq \pi$.
\end{IEEEproof}

After all the remaining \acp{ue} have been assigned to the corresponding clusters, the coordinates of the centroids are updated.
At iteration $i+1$, the new \ac{irs} configuration (centroid) of cluster $\mathcal{U}_{z, i+1}$ is computed as the average of the data points in the cluster, weighted by their achievable rate, i.e.,
\begin{equation}\label{centroidCWC}
    \bm{\Phi}^{(z)}_{i+1} = f^{-1}\left(\frac{\sum_{k \in \mathcal{U}_z}  f(\bm{\Phi}^*_k)R_k(\bm{\Phi}^*_k)}{\sum_{k \in \mathcal{U}_z}R_k(\bm{\Phi}^*_k)}\right).
\end{equation}
Also, in the case of phase shift quantization, an additional approximation step must be performed as
\begin{equation}\label{centroid_discr}
    \angle[\bm{\Phi}^{(z)}_{i+1}]_{n,n} \leftarrow
    \argmin_{\substack{\psi \in \mathcal{P}_{\theta}}}\left(\angle e^{j(\angle[\bm{\Phi}^{(z)}_{i+1}]_{n,n} - \psi)}\right), \quad \forall n.
\end{equation}
This two-step procedure is repeated until convergence, which is reached when the rate difference between two consecutive iterations is lower than the sum capacity tolerance $\mu>0$.

\begin{figure}[t]

\begin{algorithm}[H]
\caption{\acs{cwc} Algorithm}\label{alg:clustering_cwc}
\begin{algorithmic}[1]
\REQUIRE{$Z$, $\bm{H}$, $\bm{G}_k,\;  \forall k$}
\ENSURE{$\mathcal{U}_1, \ldots, \mathcal{U}_Z$, $\mathcal{I}$}
\STATE Compute $\bm{\Phi}^*_k, \forall k$ with the procedure of Algorithm~\ref{alg:ind_opt}
\STATE Sort the \acp{ue} in decreasing order of  $R_k(\bm{\Phi}^*_k)$
\STATE Select the $Z$ \acp{ue} providing the highest $R_k(\bm{\Phi}^*_k)$,  
\STATE Set $\bm{\Phi}_1^{(z)} = \bm{\Phi}^*_k$, $z = 1,\ldots, Z$ as the initial centroids.
\REPEAT
\FOR{each \ac{ue} $k$}
\STATE $z_{k, i} \gets \argmin_{\substack{z}} R_k(\bm{\Phi}^*_k) - R_k(\bm{\Phi}^{(z)}_{i})$
\ENDFOR
\FOR{each cluster $z$}
\STATE Compute $\bm{\Phi}^{(z)}_{i+1}$ as per \eqref{centroidCWC}, \eqref{centroid_discr}
\ENDFOR
\STATE$i \gets i+1$
\UNTIL{$\left|\sum_{k \in \mathcal{U}_z} R_k(\bm{\Phi}^{(z)}_{i})-\sum_{k \in \mathcal{U}_z} R_k(\bm{\Phi}^{(z)}_{i-1})\right|<\mu$}
\STATE Assign $\bm{\Phi}^{(z)}$ to all $k\in\mathcal{U}_z$.
\end{algorithmic}
\end{algorithm}

\end{figure}

The rationale behind the algorithm is that, based on the initial centroid assignment, the \acp{ue} experiencing the best channel conditions, i.e., those dominating the system sum capacity, are initially served with their optimal (individual) \ac{irs} configurations. 
Even after the adjustment of the clusters, these \acp{ue} will always get the largest weight coefficient within the cluster. The remaining \acp{ue}, instead, will be penalized by the configuration constraints, but their impact on the sum capacity will be limited.
The whole workflow of the \ac{cwc} procedure is summarized in Algorithm~\ref{alg:clustering_cwc}.

\subsection{One-Shot Capacity-Based Clustering (OSCBC)}
\label{sec:clust_oscbc}
The main drawback of \ac{cwc} presented in Section~\ref{sec:cwc} is that it requires solving problem~\eqref{cluster_assignment} at each iteration, relative to all the \acp{ue} in each cluster. 
Considering massive \ac{mimo} systems, the \ac{cwc} procedure could become exceedingly complex, as it requires the \ac{svd} computation of extremely large matrices.
Therefore, we propose another lower-complexity clustering algorithm, denoted as \ac{oscbc}.

As in \ac{cwc}, also in \ac{oscbc}:
\begin{enumerate*}[label=(\textit{\roman*})]
\item the \acp{ue} are sorted in decreasing order of achievable rate; \item the $Z$ \ac{irs} configurations of the $Z$ \acp{ue} experiencing the highest rates are chosen as initial centroids for the clusters; and \item  the remaining \acp{ue} are assigned to the closest centroid in terms of circular distance, as per \eqref{circdist}. 
\end{enumerate*}
Then, compared to \ac{cwc}, instead of recomputing the coordinates of the centroids at each iteration until convergence, the algorithm stops right after the initial association.
Therefore, with \ac{oscbc} the computed centroids are the optimal configurations relative to the $Z$ \acp{ue} achieving the highest individual rate, which provides sub-optimal (non-optimized) performance for the rest of the \acp{ue} in the clusters.

\subsection{Inverse Capacity-Weighted Clustering (ICWC)}
\label{sec:icwc}

The \ac{cwc} algorithm is designed to optimize the capacity of the \acp{ue} experiencing the best channel conditions and is unfair to the other \acp{ue} in the system, which may use sub-optimal \ac{irs} configurations. 
Therefore, we propose an additional variation of \ac{cwc}, named \ac{icwc}, with the goal of achieving higher fairness among the \acp{ue} in the system.
In \ac{icwc}, while the cluster association principle of \eqref{cluster_assignment} is preserved, the initial condition is reversed. Specifically:
\begin{enumerate*}[label=(\textit{\roman*})]
\item \acp{ue} are sorted in increasing order of achievable rate;
\item the initial configurations of the clusters $\bm{\Phi}_1^{(z)} = \bm{\Phi}^*_z$, $z = 1,\ldots, Z$, are based on the optimal configurations of the \acp{ue} with the worst channel conditions. 
\end{enumerate*}
The remaining $k>Z$ \acp{ue} are associated as in \eqref{cluster_assignment}.
Then, at iteration $i$, the \ac{irs} configuration is updated as
\begin{equation}\label{centroidICWC}
    %\bm{\Phi}^{(z)}_{i+1} = \frac{\sum_{k \in \mathcal{U}_z}  \bm{\Phi}^*_kR_k^{-1}(\bm{\Phi}^{(z)}_{i})}{\sum_{k \in \mathcal{U}_z}R_k^{-1}(\bm{\Phi}^{(z)}_{i})},
    \bm{\Phi}^{(z)}_{i+1} = f^{-1}\left(\frac{\sum_{k \in \mathcal{U}_z}  \bm{\Phi}^*_kR_k^{-1}(\bm{\Phi}^*_k)}{\sum_{k \in \mathcal{U}_z}R_k^{-1}(\bm{\Phi}^*_k)}\right),
\end{equation}
and the discretization step \eqref{centroid_discr} is performed (if needed). 
As in \ac{cwc}, convergence is achieved if the rate difference between two consecutive iterations is lower than the tolerance $\mu$.  
While \ac{icwc} obtains lower sum capacity than \ac{cwc}, it can provide significant improvements in terms of fairness, especially from the perspective of the \acp{ue} with the worst channel conditions.

\subsection{Computational Complexity}\label{complexity}

\begin{table} 
\centering
\caption{Computational complexity of distance-based vs. capacity-based clustering.}
\label{tab:clustering}
  \small
\begin{tabular}{lc}
  \toprule
Clustering algorithm & Computational complexity\\\midrule
\ac{km} (Lloyd) & $O(IZKN_{\rm I})$\\
\ac{kmed} (\acs{pam})& $O(Z^3K^2N_{\rm I})$\\
\ac{hc} & $O(K^3N_{\rm I})$ \\ \midrule
\ac{cwc}/\ac{icwc} & $O(IZKN_{\rm g}N_{\rm I}^2)$\\
\ac{oscbc} & $O(Z(K-Z)N_{\rm I})$\\
\bottomrule
\end{tabular}
\end{table}

The computational complexity is evaluated as the number of iterations required for the clustering algorithms to: \begin{enumerate*}[label=(\textit{\roman*})] 
\item obtain the optimal \ac{irs} configuration of each \ac{ue};
\item partition the \acp{ue} into disjoint subsets, or clusters, based on distance or capacity metrics; and 
\item for each cluster, find the best \ac{irs} configuration to serve the corresponding \acp{ue}.
\end{enumerate*}
%In all the algorithms, the first step is to obtain the optimal \ac{irs} configuration of each \ac{ue}, as described in Section~\ref{sec:ind_opt}. 
Specifically, at each iteration, the main source of complexity is the computation of the overall cascade channel matrix $\bm{G}_k \bm{\Phi}_k \bm{H}$, which has complexity $O\big(N_{\rm g}N_{\rm I}^2 + N_{\rm g}N_{\rm I}N_{\rm U}\big)$. %~\cite{vasudevan2017hierarchical}.
Additionally, in the case of quantized \ac{irs} phase shifts, after obtaining the optimal beamformers, the optimal phase shifts for the \ac{irs} are obtained through an exhaustive search over the set of possible phase shifts $\mathcal{P}_{\theta}$, yielding a complexity $O(2^b N_{\rm I})$.

Notice that different clustering algorithms, in general, require a different number of iterations $I$ to reach convergence, thus possibly introducing practical limitations. Moreover, the complexity introduced in each iteration depends on the clustering algorithm itself. 
In Table~\ref{tab:clustering} and in the following text we characterize the computational complexity of each of the clustering algorithms presented in Sections~\ref{sec:dist_based} and \ref{sec:cap_based}.

\vspace{2pt}\noindent\emph{Distance-based clustering.}
These algorithms do not require specific initialization.  
For \ac{km}, based on the Lloyd implementation in \cite{Kmeans}, each iteration involves calculating the distances between data points and centroids. As a result, the computational complexity is influenced by the number of iterations required for convergence, the number of data points, the number of clusters, and the dimensionality of data, resulting in an overall complexity $O(IZKN_{\rm I})$.  
\ac{kmed} can be solved with the \ac{pam} algorithm~\cite{van2003new}, so the computational complexity is $O(Z^3K^2N_{\rm I})$ due to the pairwise distance computations between data points and medoids. Finally, the computational complexity of the agglomerate \ac{hc} is primarily determined by the computation of pairwise distances among all data points, resulting in a total complexity $O(K^3N_{\rm I})$ \cite{xu2015comprehensive}.

\vspace{2pt}\noindent\emph{Capacity-based clustering.}
The complexity of the \ac{oscbc} algorithm is dominated by the centroid assignment upon initialization, which has complexity $O\big(Z(K-Z)N_{\rm I}\big)$. Instead, for the CWC and ICWC algorithms, the complexity is $O(IZKN_{\rm g}N_{\rm I}^2)$, as demonstrated in the following theorem.

\begin{theorem}
The time complexity of \ac{cwc} and \ac{icwc} scales quadratically with $N_{\rm I}$ as $O(IZKN_{\rm g}N_{\rm I}^2$).
\end{theorem}
\begin{IEEEproof}
    Capacity-based clustering requires an initialization stage where the algorithm selects the $Z$ \acp{ue} providing the highest (or lowest) $R_k\big(\bm{\Phi}^*_k\big)$, resulting in a complexity of $O(K\log K)$ due to the sorting of $K$ scalars.
    In the subsequent iterations:
    \begin{enumerate}
        \item Both \ac{cwc} and \ac{icwc} compute the rate difference between each \ac{ue} and the $Z$ centroids. The complexity of computing $R_k(\bm{\Phi}^{(z)}_{i})$ can be dominated either by the matrix multiplication in \eqref{snr_k}, or by the \ac{svd} for the single stream beamforming which require, respectively, $O\big(N_{\rm g}N_{\rm I}^2 + N_{\rm g}N_{\rm I}N_{\rm U}\big)$ and $O\big(N_{\rm g}N_{\rm U}\min(N_{\rm g},N_{\rm U})\big)$ operations for each \ac{ue} and each centroid.
        \item The computation of the centroids as per \eqref{centroidCWC}-\eqref{centroidICWC} requires $N_{\rm I}+1$ scalar operations per \ac{ue}, which has negligible complexity with respect to the rate computation.
     \end{enumerate}
    In typical \ac{irs}-assisted systems, $N_{\rm I}\gg N_{\rm g}>N_{\rm U}$. Therefore, the complexity at each iteration is dominated by the channel matrix product, and the overall algorithm complexity is $O(IZKN_{\rm g}N_{\rm I}^2)$.
\end{IEEEproof}

\section{Numerical Results}\label{sec:numerical_results}

After presenting our various simulation scenarios and evaluation metrics in~Sections~\ref{sub:sim-params} and \ref{sec:metrics}, respectively, we assess in Section~\ref{sub:results} the scheduling performance of an \ac{irs}-assisted network with practical constraints.

\subsection{Simulation Parameters}
\label{sub:sim-params}

\begin{table}[t!]
  \caption{Simulation parameters.}
  \label{Tab:parameters}
  \small
  \centering
  \begin{tabular}{lc}
    \toprule
    Parameter                  & Value \\ \midrule
    Carrier frequency		   & $28$~GHz	\\
    Total bandwidth	($B$)		   & $100$~MHz \\
    Noise power spectral density  & $-174$~dBm/Hz \\
    Number of \acp{ue} ($K$) & $100$ \\
    \ac{gnb} antenna array ($N_{\rm g}$)    & $8$H$\times8$V \\
    \ac{gnb} transmit  power	   & $33$~dBm \\
     \ac{ue} antenna array ($N_{\rm U}$)    & $2$H$\times1$V \\
    \multirow{2}{*}{\ac{irs} elements ($N_{\rm I}$)}&\{$10$H$\times20$V, $20$H$\times40$V,\\&\quad $40$H$\times80$V, $60$H$\times120$V\}\\
    Phase shift quant. bits ($b$)            & \{unquantized, $1$-bit, $2$-bits\}\\%, $5$
    \Acrshort{los} probability ($p_{\rm LoS}$)            & Eq.~\eqref{losprob}\\
    Individual rate opt. tolerance ($\nu$)            & $10^{-6}$~[bit/s/Hz]\\
    \ac{km} max. iterations ($I_{\rm max}^{\rm KM}$)            & $50$\\
    \ac{cwc}/\ac{icwc} rate tolerance ($\mu$)        & $10^{-3}$~[bit/s]\\
    \bottomrule
  \end{tabular}
\end{table}

Our simulation parameters are reported in Table~\ref{Tab:parameters}.

\vspace{2pt}\noindent\emph{Scenario.} All devices are assumed to lie on a 2D plane, and  we consider an \ac{umi} scenario, according to the 3GPP nomenclature~\cite{3gpp.38.901}, with the \ac{gnb} placed at the center. According to the 3GPP specifications, the coverage area of the \ac{gnb} is characterized by an average radius of $167$~m and is assumed to lie in the positive $x$-axis region.

We assume that $K=100$ \acp{ue} are randomly deployed according to a uniform distribution within the cell area, to be served in downlink by the \ac{gnb}, assisted by an \ac{irs} at coordinates  $(75, 100)$~m. 
The \ac{gnb} is equipped with a \ac{upa} with $8$H$\times8$V antennas (i.e., $N_{\rm g}=64$), and the \acp{ue} with \acp{ula} of $2$H$\times1$V antennas (i.e., $N_{\rm U} = 2$). 
For the \ac{irs}, if not otherwise specified, we adopt a $40$H$\times80$V reflective panel ($N_{\rm I}= 3200$).

\vspace{2pt}\noindent\emph{Channel and Frame Structure.}
The system operates at a carrier frequency of $28$~GHz (that is in the lower part of the \ac{mmwave} bands), the transmission power at the \ac{gnb} is set to $33$~dBm, the noise power spectral density at the receivers is $-174$~dBm/Hz,  and the total system bandwidth is $100$~MHz. We consider the fourth numerology of the \ac{nr} frame structure \cite{3gpp.38.211}, wherein each $10$~ms frame is split into $160$ slots. With this assumption, as already pointed out in Section~\ref{sec:system_model}, channels can be considered constant over the entire frame duration.
We consider the \ac{3gpp} TR~38.901 spatial channel model~\cite{3gpp.38.901}, which supports a wide range of frequencies, from $0.5$ to $100$ GHz (and including therefore our carrier frequency of $28$~GHz), and can be integrated with realistic beamforming models.
As such, channel matrices, and multipath fading, are computed based on the superposition of $N$ different clusters, each of which consists of $M$ rays that arrive (depart) to (from) the antenna arrays with specific angles and powers. 
Based on~\cite{3gpp.38.901}, and using the simplifications proposed in~\cite{zugno20implementation}, the generic entry $[\bm{A}]_{pq}$ of the channel matrix can then be computed as:
\begin{equation}
\label{eq:ch_model_full}
\begin{aligned}
[\bm{A}]_{pq} = \; &\gamma \sum_{n=1}^{N} \sqrt{\frac{P_{n}}{M}} \sum_{m=1}^{M} \overline{\mathbf{F}}_{r x}\left(\theta_{n, m}^{A}, \phi_{n, m}^{A}\right) \\
& \times\left[\begin{array}{cr}
e^{j \Phi_{n, m}^{\theta, \theta}} & \sqrt{K_{n, m}^{-1}} e^{j \Phi_{n, m}^{\theta, \phi}} \\
\sqrt{K_{n, m}^{-1}} e^{j \Phi_{n, m}^{\phi, \theta}} & e^{j \Phi_{n, m}^{\phi, \phi}}
\end{array}\right] \\
& \times \overline{\mathbf{F}}_{tx}\left(\theta_{n, m}^{D}, \phi_{n, m}^{D}\right) \\
& \times e^{j \overline{\mathbf{k}}_{rx, n, m}^{\rm T} \overline{\mathbf{d}}_{rx, p} e^{j \overline{\mathbf{k}}_{tx, n, m}^{\rm T} \overline{\mathbf{d}}_{tx, q}}},
\end{aligned}
\end{equation}
where $\gamma$ is the \ac{lsfc} of the considered link, which incorporates the path loss and shadowing terms. For a complete description of the remaining terms appearing in \eqref{eq:ch_model_full} we refer the interested reader to~\cite{zugno20implementation}.
Specifically, while the \ac{gnb} and the \ac{irs} can be assumed to operate in \ac{los}, the path loss between a generic \ac{ue} $k$ and the \ac{irs} is modeled based on the following channel conditions:

\noindent $\bullet$ \emph{\ac{nlos}}: \ac{ue} $k$ is in \ac{nlos} with the~\ac{irs};

\noindent $\bullet$ \emph{\ac{dlos}}: \ac{ue} $k$ is in \ac{los} with the \ac{irs};

\noindent $\bullet$ \emph{\ac{plos}}: the \ac{irs}-\ac{ue} $k$ link is in \ac{los}$\vee$\ac{nlos} with respective probabilities $p_k^{\rm LoS}(d_k)$ $\vee$ $1{-}p_k^{\rm LoS}(d_k)$, with
\begin{equation}
\label{losprob}
p_k^{\rm LoS}(d_k) = 
\begin{cases}
1 \quad &\hbox{if} \quad d_k{\leq}18, \\
\frac{18}{d_k} {+} \left(1{-}\frac{18}{d_k}\right)e^{-\frac{d_k}{36}} \quad &\hbox{if} \quad d_k{>}18,
\end{cases}
\end{equation}
where $d_k$ is the distance (in m) between the \ac{irs} and \ac{ue} $k$. In the considered \ac{umi} scenario, and based on 3GPP specifications~\cite{3gpp.38.901}, the average \ac{los} probability in~\eqref{losprob} is~$0.35$.

For each wireless link, based on the presence of the \ac{los} component, the path loss is then derived according to \cite[Table 7.4.1-1]{3gpp.38.901}, with shadowing standard deviation set to $\sigma_{\rm SF}=0$.  
For the optimal individual \ac{irs} configuration (Section~\ref{sec:ind_opt}), we set $\nu = 10^{-6}$~[bit/s/Hz].

\vspace{2pt}\noindent\emph{Clustering algorithms.}
In the following subsections, we present extensive simulation results to compare the performance of distance-based (\ac{km}, \ac{hc}, \ac{kmed}) vs. capacity-based (\ac{cwc}, \ac{oscbc}, \ac{icwc}) clustering algorithms to perform scheduling in an \ac{irs} system with reconfiguration constraints.
The \ac{km} clustering has been implemented with the Lloyd algorithm \cite{Kmeans} with a maximum number of $I_{\rm max}^{\rm KM}=50$ iterations. 
Instead, for both \ac{cwc} and \ac{icwc}, we set $\mu = 10^{-3}$~[bit/s].

As an upper bound to the system performance, we also consider an ``unclustered'' scheduling, wherein we assume that all \acp{ue} are served with their optimal \ac{irs} configuration. This scheduling clearly violates the constraint on the maximum numbers of reconfiguration per frame,  but can be regarded as the limit case when $Z = K$, i.e., all \acp{ue} belong to a cluster with cardinality one. As such, it is a suitable approach for benchmarking the performance of more practical schemes.

\subsection{Performance Metrics}
\label{sec:metrics}

The performance of the proposed clustering-based scheduling techniques is evaluated in terms of average sum capacity and fairness, as a function of the numbers of both clusters and \acp{ue}, under different channel conditions, \ac{irs} dimensions, and degrees of quantization for the phase shifts.

\vspace{2pt}\noindent\emph{Average sum capacity.} 
It is derived from \eqref{eq:sumrate} as
\begin{equation}
    \bar{C} = \frac{1}{K}\mathbb{E}\left[C(\mathcal{U}_1,\ldots,\mathcal{U}_{Z},\mathcal{I})\right],
\end{equation}
where the expectation is computed across the different channel realizations. Moreover, as each \ac{ue} is served in its specific slot, we average over the  \ac{tdma} frame length, dividing the empirical expectation by the number of \acp{ue} (slots)~$K$.  

\vspace{2pt}\noindent\emph{Fairness.}
We consider the $95$\% percentile of the achieved individual user capacity, computed as
\begin{equation}\label{C95}
    C_{95\%} = \frac{B}{K}\inf\{x: {\rm CDF}(x) \geq 0.95\},
\end{equation}
where ${\rm CDF}(\cdot)$ is the empirical cumulative distribution function  of $R_k(\bm{\Phi}^{(z)})$, $\forall k, z$. 
Notice that the $95$\% percentile of the user capacity is a practical and meaningful way to evaluate fairness, as it measures the performance of the majority of the \acp{ue}, excluding only the top 5\%.

\subsection{Scheduling Performance}
\label{sub:results}

In this section, we compare the \ac{irs} scheduling performance considering distance-based vs. capacity-based clustering, and as a function of different channel conditions, reconfiguration constraints, and degrees of quantization of the phase shifts. 

\begin{figure}[t]
    \centering
       \setlength\fwidth{0.82\columnwidth}
    \setlength\fheight{0.5\columnwidth}
     \definecolor{mycolor2}{rgb}{1,0.2,0.2}%%
\definecolor{mycolor1}{rgb}{0,0,0.9}
\definecolor{mycolor3}{rgb}{0.46600,0.67400,0.18800}%
\definecolor{mycolor7}{rgb}{0.92900,0.69400,0.12500}%
\definecolor{mycolor5}{rgb}{0.30100,0.74500,0.93300}%
\definecolor{mycolor6}{rgb}{0.49400,0.18400,0.55600}%
\definecolor{mycolor4}{rgb}{0.63500,0.07800,0.18400}%
\pgfplotsset{every tick label/.append style={font=\scriptsize}}
\begin{tikzpicture}
%%%%%%%%%%%%%%%%%%%%%%%%%%%%%%%%%%

\begin{axis}[%
    width=\fwidth,
    height=\fheight,
    at={(0\fwidth,0\fheight)},
    scale only axis,
    xlabel style={font=\footnotesize},
    ylabel style={font=\footnotesize},
    xmin=1,
    xmax=100,
    ymin=0,
    ymax=380,
    xtick={  1,  10,  20,  30,  40,  50,  60,  70,  80,  90, 100},
    xlabel={Maximum number of clusters $Z$},
    ylabel={Average sum capacity $\bar{C}$ [Mbit/s]},
    axis background/.style={fill=white},
    xmajorgrids,
    ymajorgrids,
        legend image post style={mark indices={}},
    legend style={
        /tikz/every even column/.append style={column sep=0.2cm},
        at={(0.82, 0.02)}, 
        anchor=south, 
        draw=white!80!black, 
        font=\scriptsize,
        fill opacity=0.8
        },
    legend columns=1,
]

\addplot [color=mycolor4, very thick, mark size=2.8pt, mark=square*, mark options={solid, mycolor4}]
  table[row sep=crcr]{%
1	28.1176119718827\\
10	100.861817510268\\
20	141.73705839455\\
30	184.950555262946\\
40	220.748336489813\\
50	249.049165794823\\
60	278.433938108025\\
70	298.761903651574\\
80	319.977611125998\\
90	337.45750170188\\
100	349.625593642553\\
};
\addlegendentry{KMed}

\addplot [color=mycolor5, very thick, mark size=3.0pt, mark=diamond*, mark options={solid, mycolor5}]
  table[row sep=crcr]{%
1	18.2443411157527\\
10	77.7983732080269\\
20	123.566685831694\\
30	159.811932668768\\
40	192.862674643542\\
50	221.880711483525\\
60	250.403247328401\\
70	277.018934845287\\
80	302.122691870657\\
90	325.717076087751\\
100	349.625593642553\\
};
\addlegendentry{KM}

\addplot [color=mycolor6, very thick, mark size=3.0pt, mark=pentagon*, mark options={solid, mycolor6}]
  table[row sep=crcr]{%
1	25.5007827861076\\
10	112.925237968855\\
20	166.830476313084\\
30	205.672720543757\\
40	236.778342633094\\
50	264.308292456048\\
60	285.840372512358\\
70	304.430768920583\\
80	320.543034112752\\
90	335.248581783295\\
100	349.625593642553\\
};
\addlegendentry{HC}

\addplot [color=mycolor7, very thick, mark size=3.0pt, mark=star]
  table[row sep=crcr]{%
1	349.625502666711\\
10	349.625502666711\\
20	349.625502666711\\
30	349.625502666711\\
40	349.625502666711\\
50	349.625502666711\\
60	349.625502666711\\
70	349.625502666711\\
80	349.625502666711\\
90	349.625502666711\\
100	349.625502666711\\
};
\addlegendentry{Unclustered}

\addplot [color=mycolor2, very thick, mark size=3.0pt, mark=o, mark options={solid, mycolor2}]
  table[row sep=crcr]{%
1	69.2198831219031\\
10	146.251344545728\\
20	200.447073159671\\
30	244.732697150523\\
40	276.703626326244\\
50	300.300582274394\\
60	317.284325558418\\
70	329.751795212581\\
80	338.816310164981\\
90	345.113220285055\\
100	349.625593642553\\
};
\addlegendentry{CWC}

\addplot [color=mycolor1, very thick, mark size=3.0pt, mark=x, mark options={solid, mycolor1}]
  table[row sep=crcr]{%
1	30.7889899649677\\
10	94.5687304604159\\
20	136.096494188705\\
30	171.571973470307\\
40	202.846458893204\\
50	227.85446794259\\
60	248.01792008807\\
70	269.170203404248\\
80	294.955409175835\\
90	324.880336264252\\
100	349.625593642553\\
};
\addlegendentry{ICWC}

\addplot [color=mycolor3, very thick, mark size=3.0pt, mark=triangle*, mark options={solid, rotate=180, mycolor3}]
  table[row sep=crcr]{%
1	26.388476217658\\
10	121.727908917765\\
20	186.926040773053\\
30	235.871625771207\\
40	270.739181989123\\
50	296.603964541797\\
60	315.119655587025\\
70	328.408235238309\\
80	338.086670053488\\
90	344.893216468099\\
100	349.625593642553\\
};
\addlegendentry{OSCBC}

%\addplot [color=mycolor7, dashed, very thick]

\end{axis}

\end{tikzpicture}%
    \caption{Average sum capacity as a function of the maximum number of clusters $Z$, for an unquantized $40$H$\times80$V \ac{irs}, and considering a \ac{plos} channel for the \ac{irs}-\acp{ue} links.}
    \label{fig:sumcap0bitlos}
    
\end{figure}
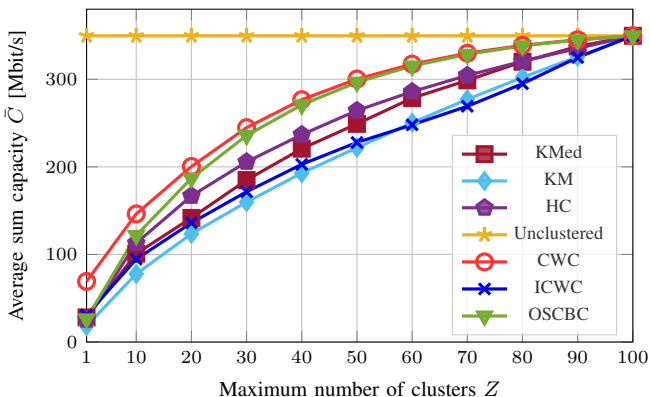

\vspace{2pt}\noindent\emph{Impact of the clustering algorithm.}
First, Fig.~\ref{fig:sumcap0bitlos} displays the average sum capacity $\bar{C}$ per slot as a function of the number of clusters $Z$, for unquantized \ac{irs} phase shifts, and considering a \ac{plos} channel for the \ac{irs}-\acp{ue} links. 
It is evident that all the scheduling policies perform better whenever $Z$ increases, and converge to the ``unclustered'' policy when $Z=K$. In fact, increasing the number of clusters corresponds to a smaller intra-cluster average distance, which eventually becomes zero when $Z=K$. 
Among the considered clustering policies, \ac{cwc} and \ac{oscbc} provide the highest sum capacity, as they are designed to maximize $\bar{C}$, and choose the \ac{irs} configurations of the \acp{ue} that achieve the highest rate. Instead, distance-based clustering achieves worse performance as it does not exploit the knowledge of the rate achievable with different \ac{irs} configurations when building the clusters.
As expected, \ac{icwc} is designed to promote fairness, thus underperforms both \ac{cwc} and \ac{oscbc} in terms of sum capacity; still, it achieves similar performance as distance-based clustering.
Finally, the gap between \ac{cwc} and \ac{oscbc} is almost negligible: this implies that a single iteration in the clustering process is enough to achieve good sum capacity, while also promoting lower computational complexity as reported in Table~\ref{tab:clustering}, which demonstrates the good scalability of the proposed techniques.

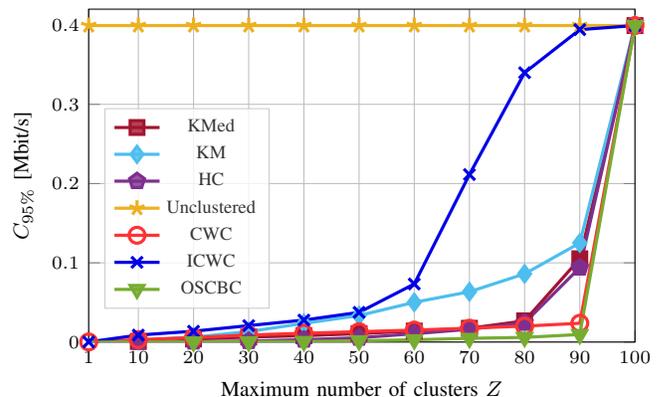
\begin{figure}[t]
    \centering
    \setlength\fwidth{0.82\columnwidth}
    \setlength\fheight{0.5\columnwidth}
    \definecolor{mycolor2}{rgb}{1,0.2,0.2}
\definecolor{mycolor1}{rgb}{0,0,0.9}
\definecolor{mycolor3}{rgb}{0.46600,0.67400,0.18800}%
\definecolor{mycolor7}{rgb}{0.92900,0.69400,0.12500}%
\definecolor{mycolor5}{rgb}{0.30100,0.74500,0.93300}%
\definecolor{mycolor6}{rgb}{0.49400,0.18400,0.55600}%
\definecolor{mycolor4}{rgb}{0.63500,0.07800,0.18400}%

\pgfplotsset{every tick label/.append style={font=\scriptsize}}
\begin{tikzpicture}

\begin{axis}[%
    width=\fwidth,
    height=\fheight,
    at={(0\fwidth,0\fheight)},
    scale only axis,
    xlabel style={font=\footnotesize},
    ylabel style={font=\footnotesize},
    xmin=1,
    xmax=100,
    ymin=0.0001,
    ymax= 0.42,
    xtick={1,  10,  20,  30,  40,  50,  60,  70,  80,  90, 100},
    xlabel={Maximum number of clusters $Z$},
    %ymode=log,
    ylabel={$C_{95\%}$ [Mbit/s]},
    axis background/.style={fill=white},
    xmajorgrids,
    ymajorgrids,
    legend style={
        /tikz/every even column/.append style={column sep=0.2cm},
        at={(0.18, 0.1)}, 
        anchor=south, 
        draw=white!80!black, 
        font=\scriptsize,   
        fill opacity=0.8
        },
    legend columns=1,
]

\addplot [color=mycolor4, very thick, mark size=2.6pt, mark=square*, mark options={solid, mycolor4}]
  table[row sep=crcr]{%
1	9.05567793095349e-05\\
10	0.000837910818601856\\
20	0.00392855662136112\\
30	0.00638085608243083\\
40	0.0080831904650645\\
50	0.0106925702195818\\
60	0.0139666990027101\\
70	0.0174186473196289\\
80	0.0268293291425108\\
90	0.105119848182379\\
100	0.39935086872942\\
};
\addlegendentry{KMed}

\addplot [color=mycolor5, very thick, mark size=3.0pt, mark=diamond*, mark options={solid, mycolor5}]
  table[row sep=crcr]{%
1	6.43043259014252e-05\\
10	0.00204437072674061\\
20	0.00624249618170619\\
30	0.0132722920611753\\
40	0.0236523001448433\\
50	0.033514085573131\\
60	0.0501833715747626\\
70	0.0635192697665196\\
80	0.0861044601235812\\
90	0.125070416781253\\
100	0.39935086872942\\
};
\addlegendentry{KM}

\addplot [color=mycolor6, very thick, mark size=3.0pt, mark=pentagon*, mark options={solid, mycolor6}]
  table[row sep=crcr]{%
1	5.08142594851443e-05\\
10	0.000322797610389904\\
20	0.000707754344589662\\
30	0.00131691241135746\\
40	0.00304059379386011\\
50	0.00549901671162177\\
60	0.0105967521074395\\
70	0.0161713287105324\\
80	0.0243123351508351\\
90	0.0939923850239182\\
100	0.39935086872942\\
};
\addlegendentry{HC}

\addplot [color=mycolor7, very thick, mark size=3.0pt, mark=star]
  table[row sep=crcr]{%
1	0.399349026377281\\
10	0.399349026377281\\
20	0.399349026377281\\
30	0.399349026377281\\
40	0.399349026377281\\
50	0.399349026377281\\
60	0.399349026377281\\
70	0.399349026377281\\
80	0.399349026377281\\
90	0.399349026377281\\
100	0.399349026377281\\
};
\addlegendentry{Unclustered}

\addplot [color=mycolor2, very thick, mark size=3.0pt, mark=o, mark options={solid, mycolor2}]
  table[row sep=crcr]{%
1	0.000201347131794128\\
10	0.00312580402520538\\
20	0.00573819025062837\\
30	0.00895436056303841\\
40	0.011225030668531\\
50	0.0131333117023435\\
60	0.0151559423415213\\
70	0.0175307211254688\\
80	0.0201655844663867\\
90	0.023766885578881\\
100	0.39935086872942\\
};
\addlegendentry{CWC}

\addplot [color=mycolor1, very thick, mark size=3.0pt, mark=x, mark options={solid, mycolor1}]
  table[row sep=crcr]{%
1	0.00021353977713999\\
10	0.00885511846811477\\
20	0.013744379545304\\
30	0.0208065402208952\\
40	0.0277891151667394\\
50	0.0375180036653679\\
60	0.0733218322465392\\
70	0.21139661183481\\
80	0.339879349221883\\
90	0.394219418443468\\
100	0.39935086872942\\
};
\addlegendentry{ICWC}

\addplot [color=mycolor3, very thick, mark size=3.0pt, mark=triangle*, mark options={solid, rotate=180, mycolor3}]
  table[row sep=crcr]{%
1	2.62521873506631e-05\\
10	6.81291054156293e-05\\
20	0.000284919488032969\\
30	0.000440668563758382\\
40	0.000786034745828592\\
50	0.00169163539622302\\
60	0.00294921232516224\\
70	0.00475839240695278\\
80	0.00596724043372362\\
90	0.00958303324174252\\
100	0.39935086872942\\
};
\addlegendentry{OSCBC}

%\addplot [color=mycolor7, dashed, very thick]

\end{axis}

\end{tikzpicture}
    \caption{$95$\% percentile of the user capacity as a function of the maximum number of clusters   $Z$, for an unquantized $40$H$\times80$V \ac{irs}, and considering a \ac{plos} channel for the \ac{irs}-\acp{ue} links.}
    
    \label{fig:quantile0bitsplos}
\end{figure}

Fig.~\ref{fig:quantile0bitsplos} compares the fairness performance of the different clustering algorithms, measured as the $95$\% percentile of the average sum capacity $C_{95\%}$, as a function of the maximum number of clusters $Z$ in \ac{plos} conditions. 
Our results identify \ac{icwc} as the best clustering approach in terms of fairness, which comes at the cost of a lower sum capacity,  as shown in Fig.~\ref{fig:sumcap0bitlos}. 
Therefore, there exists a trade-off between the achievable sum capacity and fairness. 
We also observe that \ac{oscbc} achieves very low fairness, as the \acp{ue} with worst channel conditions are forced to aggregate to the strongest \acp{ue}, thus via a sub-optimal \ac{irs} configuration. On the other hand, we see that \ac{cwc} is more than acceptable in terms of fairness, and achieves comparable performance than most of the distance-based clustering algorithms.
Finally, notice that $C_{95\%}$ increases as $Z$ increases, and eventually approaches the ``unclustered'' baseline for $Z=K$. This is due to the fact that the \ac{los} probability in the \ac{plos} scenario increases with the number of clusters, i.e., as the inter-cluster distance becomes smaller, which permits to experience better channel conditions, thus a higher capacity, even for the worst \acp{ue}.

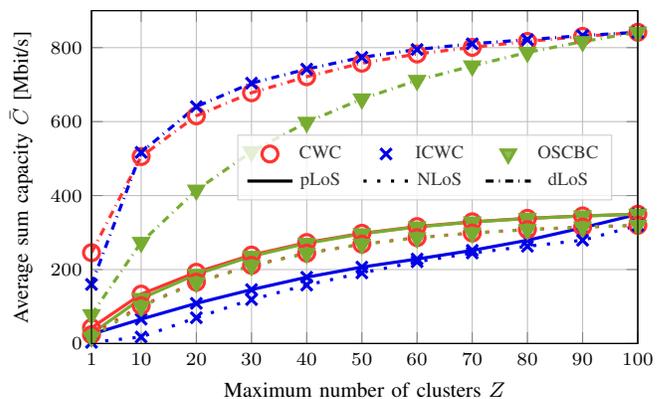
\begin{figure}[t]
    \centering
    \setlength\fwidth{0.82\columnwidth}
    \setlength\fheight{0.5\columnwidth}
    \definecolor{mycolor2}{rgb}{1,0.2,0.2}%%ICWC
\definecolor{mycolor7}{rgb}{0.92900,0.69400,0.12500}
\definecolor{mycolor1}{rgb}{0,0,0.9}
\definecolor{mycolor3}{rgb}{0.46600,0.67400,0.18800}
\pgfplotsset{every tick label/.append style={font=\scriptsize}}
\begin{tikzpicture}

\begin{axis}[%
    width=\fwidth,
    height=\fheight,
    at={(0\fwidth,0\fheight)},
    scale only axis,
        xlabel style={font=\footnotesize},
    ylabel style={font=\footnotesize},
    xmin=1,
    xmax=100,
    xtick={1,  10,  20,  30,  40,  50,  60,  70,  80,  90, 100},
    xlabel={Maximum number of clusters $Z$},
    ymin=0,
    ymax=900,
    ylabel={Average sum capacity $\bar{C}$ [Mbit/s]},
    axis background/.style={fill=white},
    xmajorgrids,
    ymajorgrids,
    %legend style={at={(0.65,0.02)}, anchor=south east, legend columns=2, legend cell align=left, align=left, font=\footnotesize, draw=white!15!black} %at={(0.93, 1.01)} to put it on top
        legend style={
        /tikz/every even column/.append style={column sep=0.2cm},
        at={(0.5, 1.02)}, 
        anchor=south, 
        draw=white!80!black, 
        font=\footnotesize
        },
    legend columns=3,
]

\addplot [color=mycolor2, very thick, mark size=3.0pt, mark=o, mark options={solid, mycolor2}]
  table[row sep=crcr]{%
1	42.6008062283675\\
10	133.455275985272\\
20	192.53100140295\\
30	239.069451743273\\
40	272.940118487222\\
50	297.976899669323\\
60	316.149381210092\\
70	329.094029137403\\
80	338.441817897034\\
90	344.996763882116\\
100	349.625593642553\\
};
\addlegendentry{CWC pLoS}

\addplot [color=mycolor1, very thick, mark size=3.0pt, mark=x, mark options={solid, mycolor1}]
  table[row sep=crcr]{%
1	24.9723982410139\\
10	65.8145358804775\\
20	108.54033550891\\
30	145.690036789427\\
40	179.197974806917\\
50	206.619165643715\\
60	228.386493499908\\
70	251.827830535404\\
80	279.15626941861\\
90	312.926085538803\\
100	349.625593642553\\
};
\addlegendentry{ICWC pLoS}

\addplot [color=mycolor3, very thick, mark size=3.0pt, mark=triangle*, mark options={solid, rotate=180}]
  table[row sep=crcr]{%
1	26.388476217658\\
10	121.727908917765\\
20	186.926040773053\\
30	235.871625771207\\
40	270.739181989123\\
50	296.603964541797\\
60	315.119655587025\\
70	328.408235238309\\
80	338.086670053488\\
90	344.893216468099\\
100	349.625593642553\\
};
\addlegendentry{OSCBC pLoS}

% \addplot [color=mycolor7, very thick, mark size=3.0pt, mark=star]
% table[row sep=crcr]{%
% 1	349.625593642553\\
% 10	349.625593642553\\
% 20	349.625593642553\\
% 30	349.625593642553\\
% 40	349.625593642553\\
% 50	349.625593642553\\
% 60	349.625593642553\\
% 70	349.625593642553\\
% 80	349.625593642553\\
% 90	349.625593642553\\
% 100	349.625593642553\\
% };
% \addlegendentry{Unclustered NLoS}

\addplot [color=mycolor2, loosely dotted, very thick, mark size=3.0pt, mark=o, mark options={solid}]
  table[row sep=crcr]{%
1	23.9593499753299\\
10	102.137688982149\\
20	165.847823326129\\
30	211.332549717504\\
40	244.34284555178\\
50	268.726542036801\\
60	286.206729878088\\
70	298.861833092044\\
80	308.163135964289\\
90	314.846491506187\\
100	319.556593125526\\
};
\addlegendentry{CWC NLoS}

\addplot [color=mycolor1, loosely dotted, very thick, mark size=3.0pt, mark=x, mark options={solid}]
  table[row sep=crcr]{%
1	3.09474410129819\\
10	17.5767002186557\\
20	69.7113378612216\\
30	118.800128113958\\
40	158.015850583673\\
50	190.893509083211\\
60	221.033239679313\\
70	244.690885409656\\
80	262.891391449184\\
90	279.240889892842\\
100	319.286618387674\\
};
\addlegendentry{ICWC NLoS}

\addplot [color=mycolor3, loosely dotted, very thick, mark size=3.0pt, mark=triangle*, mark options={solid,rotate=180}]
  table[row sep=crcr]{%
1	17.1231572037836\\
10	100.26327785691\\
20	164.494680485333\\
30	210.273242236461\\
40	243.648553079664\\
50	268.120074236045\\
60	285.804111784643\\
70	298.615600299055\\
80	308.006754232901\\
90	314.810586513323\\
100	319.596150399259\\
};
\addlegendentry{OSCBC NLoS}

% \addplot [color=mycolor7, loosely dotted, very thick, mark size=3.0pt, mark=star, mark options={solid}]
% table[row sep=crcr]{%
% 1	319.596150399259\\
% 10	319.596150399259\\
% 20	319.596150399259\\
% 30	319.596150399259\\
% 40	319.596150399259\\
% 50	319.596150399259\\
% 60	319.596150399259\\
% 70	319.596150399259\\
% 80	319.596150399259\\
% 90	319.596150399259\\
% 100	319.596150399259\\
% };
% \addlegendentry{Unclustered NLoS}

\addplot [color=mycolor2, dashdotted, very thick, mark size=3.0pt, mark=o, mark options={solid}]
  table[row sep=crcr]{%
1	245.575247920652\\
10	505.576001687911\\
20	615.646718047789\\
30	678.005019204302\\
40	721.415976205122\\
50	758.005919643906\\
60	782.82554107869\\
70	800.729608627994\\
80	816.687523407408\\
90	830.424727362428\\
100	841.603020097635\\
};
\addlegendentry{CWC LoS}

\addplot [color=mycolor1, dashdotted, very thick, mark size=3.0pt, mark=x, mark options={solid}]
  table[row sep=crcr]{%
1	159.681267871958\\
10	516.80948143988\\
20	640.589179704565\\
30	704.006653149399\\
40	741.92346296552\\
50	773.753198056541\\
60	794.99893577607\\
70	810.439215609657\\
80	821.541372542533\\
90	832.571493201843\\
100	841.603020097635\\
};
\addlegendentry{ICWC LoS}

\addplot [color=mycolor3, dashdotted, very thick, mark size=3.0pt, mark=triangle*, mark options={solid,rotate=180}]
  table[row sep=crcr]{%
1	77.9292502635754\\
10	273.39800608375\\
20	414.621582406316\\
30	518.68482471127\\
40	597.99838951715\\
50	661.22081807051\\
60	711.397310227455\\
70	750.535897300065\\
80	787.104621104615\\
90	816.265965456586\\
100	841.603020097635\\
};
\addlegendentry{OSCBC LoS}

% \addplot [color=mycolor7, dashdotted, very thick, mark size=3.0pt, mark=star]
% table[row sep=crcr]{%
% 1	841.603020097635\\
% 10	841.603020097635\\
% 20	841.603020097635\\
% 30	841.603020097635\\
% 40	841.603020097635\\
% 50	841.603020097635\\
% 60	841.603020097635\\
% 70	841.603020097635\\
% 80	841.603020097635\\
% 90	841.603020097635\\
% 100	841.603020097635\\
% };
% \addlegendentry{Unclustered LoS}

\legend{}
\end{axis}

\begin{axis}[%
    width=\fwidth,
    height=\fheight,
    at={(0\fwidth,0\fheight)},
    scale only axis,
    xmin=1,
    xmax=100,
    xtick={},
    ytick={},
    xticklabels={{}, {}, {},{}},
    yticklabels={},
    xtick style = {draw=none},
    ytick style = {draw=none},
    ymin=0,
    ymax=850,
        legend style={
        /tikz/every even column/.append style={column sep=0.2cm},
        at={(0.61, 0.43)}, 
        anchor=south, 
        draw=white!80!black, 
        font=\scriptsize,
        fill opacity=0.8
        },
    legend columns=3,
]
\addplot[color=mycolor2, only marks, very thick, mark size=3.0pt, mark=o, mark options={solid}] table[row sep=crcr]{%
1	-5 \\
};
\addlegendentry{CWC}
\addplot[color=mycolor1, only marks, very thick, mark size=3.0pt, mark=x, mark options={solid}] table[row sep=crcr] {%
1	-5 \\
};
\addlegendentry{ICWC}
\addplot[color=mycolor3, only marks, very thick, mark size=3.0pt, mark=triangle*, mark options={solid, rotate=180}] table[row sep=crcr] {%
1	-5 \\
};
\addlegendentry{OSCBC}
% \addplot[color=mycolor7, only marks, very thick, mark size=3.0pt, mark=star, mark options={solid}] table[row sep=crcr] {%
% 1	-5 \\
% };
% \addlegendentry{Unclustered}
\addplot [color=black, solid, very thick]
  table[row sep=crcr]{%
1	-5\\
};
\addlegendentry{pLoS}
\addplot [color=black, loosely dotted, very thick]
  table[row sep=crcr]{%
1	-5\\
};
\addlegendentry{NLoS}
\addplot [color=black, dashdotted, very thick]
  table[row sep=crcr]{%
1	-5\\
};
\addlegendentry{dLoS}
\end{axis}

\end{tikzpicture}%
    \caption{Average sum capacity as a function of the maximum number of clusters $Z$, for $N_{\rm I}=3200$, unquantized phase shifts, and for different channel conditions.}
    \label{fig:sumcaplos}
\end{figure}

\begin{figure}[t]
    \centering
       \setlength\fwidth{0.82\columnwidth}
    \setlength\fheight{0.5\columnwidth}
    \input{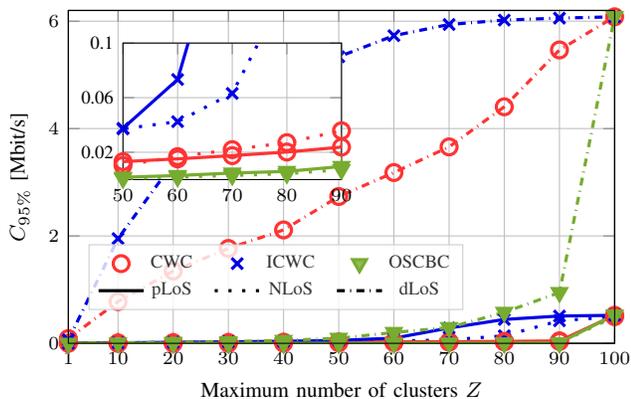}  
    \caption{$95$\% percentile of the user capacity as a function of the maximum number of clusters  $Z$,  for $N_{\rm I}=3200$, unquantized phase shifts, and for different channel conditions. The zoom inside the figure is for $50\leq Z \leq 90$.}  
    \label{fig:quantitelos}
\end{figure}

\vspace{2pt}\noindent\emph{Impact of the channel.}
From the above results, we concluded that distance-based clustering provides lower sum capacity and fairness compared to capacity-based scheduling, so the rest of our simulation campaign has been focused on the latter. 
Figs.~\ref{fig:sumcaplos} and \ref{fig:quantitelos} display the average sum capacity and the $95$\% percentile, respectively, for \ac{cwc}, \ac{icwc}, and \ac{oscbc} in different channel conditions.
First, we observe that in the \ac{dlos} scenario, where \acp{ue} are in \ac{los} with the \ac{irs}, the sum capacity is up to 2.6 (2.4) times higher than in the \ac{nlos} (\ac{plos}) scenario for $Z=K$.
This is mainly due to the fact that \ac{nlos} links experience
\begin{enumerate*}[label=(\textit{\roman*})]
  \item a higher path loss, and
  \item the lack of a dominant multipath component, thus of a clear steering direction for the \ac{irs} beam, which deteriorates the link quality. \end{enumerate*} 
In particular, in the \ac{plos} scenario the \ac{los} probability decreases exponentially with the distance, therefore, the \acp{ue} that are far from the \ac{irs} typically operate in \ac{nlos}.
For similar reasons, both \ac{cwc} and \ac{icwc} in the \ac{dlos} scenario start to reach stability in terms of capacity with a relatively lower number of clusters than in the \ac{plos} and \ac{nlos} scenarios. 

As expected, \ac{oscbc} performs worse than its competitors, and the gap is even more significant in the \ac{dlos} scenario (around $-30\%$ in terms of sum capacity).
The bad performance of \ac{oscbc} compared to \ac{cwc} and \ac{icwc} is confirmed also in terms of fairness, as illustrated in Fig. \ref{fig:quantitelos} (see, in particular, the zoom for $50\leq Z \leq 90$).

Finally, even though \ac{icwc} is not explicitly designed to maximize the sum capacity, it shows similar performance (if not even slightly better) as \ac{cwc} in the \ac{dlos} scenario.
The rationale behind this behavior is not clear and deserves more investigation. Most likely, it is related to the fact that, in the \ac{dlos} scenario, all \acp{ue} have similar channel conditions, which permits \ac{icwc} to choose, on average, a good \ac{irs} configuration even among the worst \acp{ue} in the clusters.

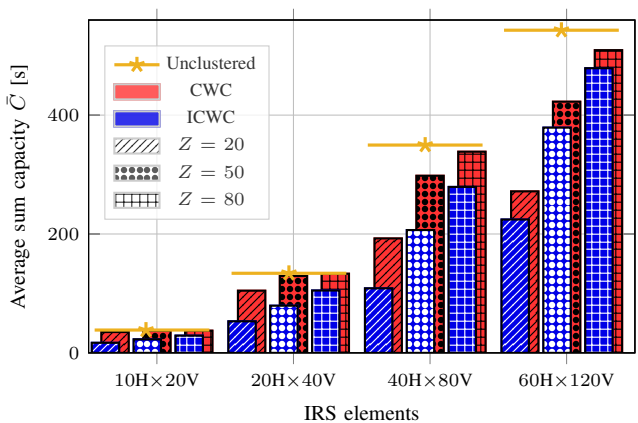
\begin{figure}[t]
    \centering
    \setlength\fwidth{0.82\columnwidth}
    \setlength\fheight{0.5\columnwidth}
    \definecolor{mycolor1}{rgb}{1,0.2,0.2}%%ICWC
\definecolor{mycolor2}{rgb}{0,0,0.9}%%CWC
\definecolor{mycolor7}{rgb}{0.92900,0.69400,0.12500}

\pgfplotsset{every tick label/.append style={font=\scriptsize}}
\begin{tikzpicture}

\begin{axis}[%
    width=\fwidth,
    height=\fheight,
    at={(0\fwidth,0\fheight)},
scale only axis,
bar shift auto,
xmin=0.5,
xmax=4.5,
ybar=5.5pt,
xtick={1,2,3,4},
xticklabels={{$10$H$\times20$V},{$20$H$\times40$V},{$40$H$\times80$V},{$60$H$\times120$V}},
ymin=0,
ymax=560,
    xlabel style={font=\footnotesize},
    ylabel style={font=\footnotesize},
ylabel={Average sum capacity $\bar{C}$ [s]},
xlabel={IRS elements},
axis background/.style={fill=white},
xmajorgrids,
ymajorgrids,
legend style={at={(0.03,0.97)}, anchor=north west, legend cell align=left, font=\footnotesize, align=left, draw=none},
axis line style={draw=none}
]
\addplot[ybar, bar width=0.2, preaction={fill, mycolor1}, thick, pattern={north east lines}, pattern color=black, draw=black, area legend] table[row sep=crcr] {%
1	34.9976\\
2	104.5981\\
3	192.5310\\
4	271.7077\\
};
\addplot[forget plot, color=white!15!black] table[row sep=crcr] {%
0.511111111111111	0\\
4.48888888888889	0\\
};
\addlegendentry{CWC $Z=20$}

\addplot[ybar, bar width=0.2, preaction={fill, mycolor1}, thick, pattern={Dots[radius=1.2]}, pattern color=black, draw=black, area legend] table[row sep=crcr] {%
1	37.1648\\
2	129.2787\\
3	297.9769\\
4	422.6018\\
};
\addplot[forget plot, color=white!15!black] table[row sep=crcr] {%
0.511111111111111	0\\
4.48888888888889	0\\
};
\addlegendentry{CWC $Z=50$}

\addplot[ybar, bar width=0.2, preaction={fill, mycolor1}, thick, pattern={grid}, pattern color=black, area legend] table[row sep=crcr] {%
1	37.3609 \\
2	133.1811\\
3	338.4418\\
4	509.2955\\
};
\addplot[forget plot, color=white!15!black] table[row sep=crcr] {%
0.511111111111111	0\\
4.48888888888889	0\\
};
\addlegendentry{CWC $Z=80$}
\legend{}
\end{axis}

\begin{axis}[%
    width=\fwidth,
    height=\fheight,
    at={(0\fwidth,0\fheight)},
scale only axis,
bar shift auto,
ybar=5.5pt,
xmin=0.57,
xmax=4.57,
xtick={1,2,3,4},
ytick={},
xticklabels={{}, {}, {},{}},
yticklabels={},
ymin=0,
ymax=560,
xtick style = {draw=none},
%ylabel={$\bar{C}$ [Mbit/time slot]},
%axis background/.style={fill=white},
%xmajorgrids,
%ymajorgrids,
legend style={at={(0.38,0.97)}, anchor=north west, legend cell align=left, font=\footnotesize, align=left, draw=white!15!black}
]
\addplot[ybar, bar width=0.2, preaction={fill, mycolor2}, thick, pattern={north east lines}, pattern color=white, draw=black, area legend] table[row sep=crcr] {%
1	16.9103 \\
2	53.0505\\
3	108.5403\\
4   224.3472\\
};
\addplot[forget plot, color=white!15!black] table[row sep=crcr] {%
0.511111111111111	0\\
4.48888888888889	0\\
};
\addlegendentry{ICWC $Z=20$}

\addplot[ybar, bar width=0.2, preaction={fill, mycolor2}, thick, pattern={Dots[radius=1.2]}, pattern color=white, area legend] table[row sep=crcr] {%
1	22.7761\\
2	79.3228\\
3	206.6192\\
4   378.9325\\
};
\addplot[forget plot, color=white!15!black] table[row sep=crcr] {%
0.511111111111111	0\\
4.48888888888889	0\\
};
\addlegendentry{ICWC $Z=50$}

\addplot[ybar, bar width=0.2, preaction={fill, mycolor2}, thick, pattern={grid}, pattern color=white, area legend] table[row sep=crcr] {%
1	29.0170\\
2	104.9147\\
3	279.1563\\
4   478.8952\\
};
\addplot[forget plot, color=white!15!black] table[row sep=crcr] {%
0.511111111111111	0\\
4.48888888888889	0\\
};
\addlegendentry{ICWC $Z=80$}
\legend{}

\end{axis}

%%%%%%%%%%%%%%%%%%%% UNCLUSTERED

\begin{axis}[%
    width=\fwidth,
    height=\fheight,
    at={(0\fwidth,0\fheight)},
scale only axis,
xmin=0.5,
xmax=4.57,
ytick={},
xticklabels={{}, {}, {},{}},
yticklabels={},
ymin=0,
ymax=560,
xtick style = {draw=none},
]
\addplot [color=mycolor7, very thick]
  table[row sep=crcr]{%
0.545	38.3936\\
1.4	38.3936\\
};
\addlegendentry{Unclustered}
\addplot [color=mycolor7, only marks, very thick, mark size=3.0pt, mark=star]
  table[row sep=crcr]{%
0.9275 38.3936\\
};

\addplot [color=mycolor7, very thick]
  table[row sep=crcr]{%
1.565	133.895\\
2.42	133.895\\
};
\addlegendentry{Unclustered}
\addplot [color=mycolor7, only marks, very thick, mark size=3.0pt, mark=star]
  table[row sep=crcr]{%
1.9925 133.895\\
};

\addplot [color=mycolor7, very thick]
  table[row sep=crcr]{%
2.58	349.625502666711\\
3.435	349.625502666711\\
};
\addlegendentry{Unclustered}
\addplot [color=mycolor7, only marks, very thick, mark size=3.0pt, mark=star]
  table[row sep=crcr]{%
3.0075  349.625502666711\\
};

\addplot [color=mycolor7, very thick]
  table[row sep=crcr]{%
3.595	542.8657\\
4.45	542.8657\\
};
\addlegendentry{Unclustered}
\addplot [color=mycolor7, only marks, very thick, mark size=3.0pt, mark=star]
  table[row sep=crcr]{%
4.0225  542.8657\\
};

\legend{}
\end{axis}

\begin{axis}[%
    width=\fwidth,
    height=\fheight,
    at={(0\fwidth,0\fheight)},
    scale only axis,
    xmin=0.511111111111111,
    xmax=4.48888888888889,
    xtick={},
    ytick={},
    xticklabels={{}, {}, {},{}},
    yticklabels={},
    ymin=0,
    ymax=560,
    xtick style = {draw=none},
    ytick style = {draw=none},
    legend style={
            /tikz/every even column/.append style={column sep=0.2cm},
            at={(0.18, 0.4)}, 
            anchor=south, 
            draw=white!80!black, 
            font=\scriptsize,
            fill opacity=0.8
            },
        legend columns=1
]

\addlegendimage{color = mycolor7, mark=star, very thick,  mark size=3.0pt}
\addlegendentry{Unclustered}
\addlegendimage{bar width=0.2, preaction={fill, mycolor1}, thick, area legend}
\addlegendentry{CWC}
\addlegendimage{bar width=0.2, preaction={fill, mycolor2}, thick, area legend}
\addlegendentry{ICWC}
\addlegendimage{bar width=0.2, preaction={fill, white}, thick, pattern={north east lines}, pattern color=black, area legend}
\addlegendentry{$Z=20$}
\addlegendimage{bar width=0.2, preaction={fill, white}, thick, pattern={Dots[radius=1.2]}, pattern color=black, area legend}
\addlegendentry{$Z=50$}
\addlegendimage{bar width=0.2, preaction={fill, white}, thick, pattern={grid}, pattern color=black, area legend}
\addlegendentry{$Z=80$}

\end{axis}

\end{tikzpicture}% 
    \caption{Average sum capacity for \ac{cwc} and \ac{icwc} as a function of the number of reflecting elements at the \ac{irs}, for unquantized phase shifts, and considering a \ac{plos} channel for the \ac{irs}-\acp{ue} links.}
    \label{fig:sumcap_vs_size}
\end{figure}

\begin{figure}[t]
    \centering
    \setlength\fwidth{0.82\columnwidth}
    \setlength\fheight{0.5\columnwidth}
    \input{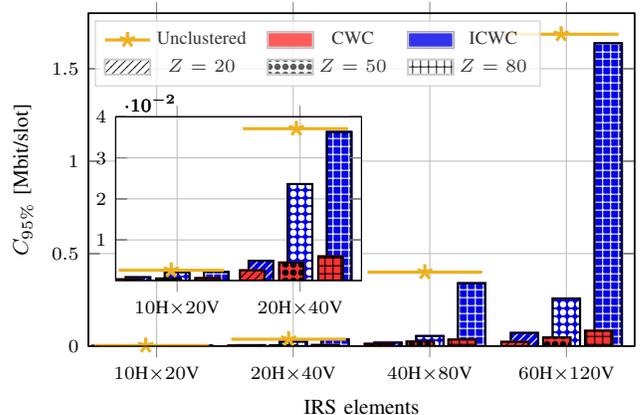}  
    \caption{$95$\% percentile of the user capacity for \ac{cwc} and \ac{icwc} as a function of the number of reflecting elements at the \ac{irs}, for unquantized phase shifts, and considering a \ac{plos} channel for the \ac{irs}-\acp{ue} links. The zoom inside the figure is for $N_{\rm I} = 200$ and $N_{\rm I} = 800$.}
    \label{fig:quantile_vs_size}
\end{figure}

\vspace{2pt}\noindent\emph{Impact of the \ac{irs} configuration.}
Figs. \ref{fig:sumcap_vs_size} and \ref{fig:quantile_vs_size} show the impact of the number of \ac{irs} radiating elements on the system performance when considering the \ac{cwc} and \ac{icwc} clustering algorithms. 
As expected, both fairness (measured in terms of the $95$\% percentile of the average sum capacity) and sum capacity increase as the \ac{irs} is larger and operates with more reflecting elements, regardless of the number of clusters.
For example, we observe that \ac{cwc} is able to approach the optimal sum capacity with as few as $20$ clusters for small-sized \ac{irs}, i.e., with $10$H$\times20$V or $20$H$\times40$V arrays. The same trends are shown also in Fig.~\ref{fig:quantile_vs_size} in terms of fairness.
Still, notice that $\bar{C}$ is below 100~Mbps, which is not compatible with the requirement of most 5G applications when the \ac{irs} is made of fewer than 200 elements, which justifies the use of larger \ac{irs} panels~\cite{pagin2022end}. 

Nevertheless, we still observe that the number of reflecting elements has an impact on the number of clusters that are needed to provide maximum performance. 
Indeed, the number of possible \ac{irs} configurations increases as we consider larger \ac{irs} antennas. In turn, this decreases the likelihood of \acp{ue} having the same (or similar) ideal configurations, and therefore, it increases the probability of being associated with increasingly sub-optimal centroids if the number of clusters is small. However, if the number of phase shifters is large, the sub-optimality is mitigated by the increasing number of reconfigurations. 

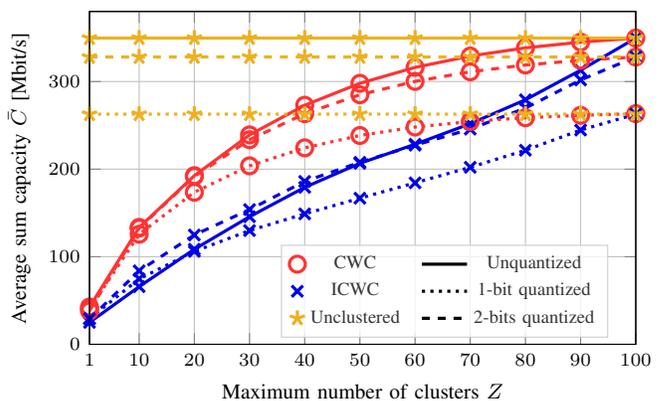
\begin{figure}[t]
    \centering
    \setlength\fwidth{0.82\columnwidth}
    \setlength\fheight{0.5\columnwidth}
    \definecolor{mycolor2}{rgb}{1,0.2,0.2}%%ICWC
%\definecolor{mycolor1}{rgb}{0,0,0}%%CWC
\definecolor{mycolor1}{rgb}{0,0,0.9}
\definecolor{mycolor7}{rgb}{0.92900,0.69400,0.12500}
\pgfplotsset{every tick label/.append style={font=\scriptsize}}
\begin{tikzpicture}

\begin{axis}[%
    width=\fwidth,
    height=\fheight,
    at={(0\fwidth,0\fheight)},
    scale only axis,
        xlabel style={font=\footnotesize},
    ylabel style={font=\footnotesize},
    xmin=1,
    xmax=100,
    xtick={1,  10,  20,  30,  40,  50,  60,  70,  80,  90, 100},
    xlabel={Maximum number of clusters $Z$},
    ymin=0,
    ymax=380,
    ylabel={Average sum capacity $\bar{C}$ [Mbit/s]},
    axis background/.style={fill=white},
    xmajorgrids,
    ymajorgrids,
    legend style={at={(0.65,0.02)}, anchor=south east, legend columns=2, legend cell align=left, align=left, font=\footnotesize, draw=white!15!black} %at={(0.93, 1.01)} to put it on top
]

\addplot [color=mycolor1, very thick, mark size=3.0pt, mark=x, mark options={solid, mycolor1}]
  table[row sep=crcr]{%
1	24.9723982410139\\
10	65.8145358804775\\
20	108.54033550891\\
30	145.690036789427\\
40	179.197974806917\\
50	206.619165643715\\
60	228.386493499908\\
70	251.827830535404\\
80	279.15626941861\\
90	312.926085538803\\
100	349.625593642553\\
};
\addlegendentry{ICWC Unquantized}

\addplot [color=mycolor2, very thick, mark size=3.0pt, mark=o, mark options={solid, mycolor2}]
  table[row sep=crcr]{%
1	42.6008062283675\\
10	133.455275985272\\
20	192.53100140295\\
30	239.069451743273\\
40	272.940118487222\\
50	297.976899669323\\
60	316.149381210092\\
70	329.094029137403\\
80	338.441817897034\\
90	344.996763882116\\
100	349.625593642553\\
};
\addlegendentry{CWC Unquantized}

\addplot [color=mycolor7, very thick, very thick, mark size=3.0pt, mark=star, mark options={solid}]
  table[row sep=crcr]{%
1	349.625593642553\\
10	349.625593642553\\
20	349.625593642553\\
30	349.625593642553\\
40	349.625593642553\\
50	349.625593642553\\
60	349.625593642553\\
70	349.625593642553\\
80  349.625593642553\\
90	349.625593642553\\
100	349.625593642553\\
};
\addlegendentry{Unclustered $b = 1$}

\addplot [color=mycolor2, dashed, very thick, mark size=3.0pt, mark=o, mark options={solid}]
  table[row sep=crcr]{%
1	38.1344267299891\\
10	132.663573598247\\
20	191.662759950297\\
30	233.568284748996\\
40	263.294436492775\\
50	284.913653622265\\
60	300.173959979977\\
70	311.116287568021\\
80	318.899256859483\\
90	324.228342378166\\
100	327.999688618062\\
};
\addlegendentry{CWC $b = 2$}

\addplot [color=mycolor1, dashed, very thick, mark size=3.0pt, mark=x, mark options={solid}]
  table[row sep=crcr]{%
1	28.8823626297359\\
10	83.9151546787062\\
20	124.811477836246\\
30	153.844261163587\\
40	185.964619156053\\
50	207.781980866694\\
60	227.325166845796\\
70	245.659216869642\\
80	269.730734492892\\
90	301.809505857585\\
100	328.040872280513\\
};
\addlegendentry{ICWC $b =2$}

\addplot [color=mycolor7, dashed, very thick, very thick, mark size=3.0pt, mark=star, mark options={solid}]
  table[row sep=crcr]{%
1	328.040872280513\\
10	328.040872280513\\
20	328.040872280513\\
30	328.040872280513\\
40	328.040872280513\\
50	328.040872280513\\
60	328.040872280513\\
70	328.040872280513\\
80  328.040872280513\\
90	328.040872280513\\
100	328.040872280513\\
};
\addlegendentry{Unclustered $b = 1$}

\addplot [color=mycolor2, dotted, very thick, mark size=3.0pt, mark=o, mark options={solid}]
  table[row sep=crcr]{%
1	40.4488714222318\\
10	125.628828473115\\
20	173.623440116182\\
30	203.756743321695\\
40	224.307128412182\\
50	238.389354565415\\
60	248.040607397182\\
70	254.359383400904\\
80	258.644019904079\\
90	261.449592581773\\
100	263.33356512743\\
};
\addlegendentry{CWC $b = 1$}

\addplot [color=mycolor1, dotted,very thick, mark size=3.0pt, mark=x, mark options={solid}]
  table[row sep=crcr]{%
1	29.2928867803151\\
10	74.6866938503446\\
20	106.080914791654\\
30	129.871442646802\\
40	148.901748347542\\
50	166.811163181046\\
60	184.338823567086\\
70	202.106211662858\\
80	221.54488808895\\
90	244.561542345882\\
100	262.774393414388\\
};
\addlegendentry{ICWC $b =1$}

\addplot [color=mycolor7, dotted, very thick, very thick, mark size=3.0pt, mark=star, mark options={solid}]
  table[row sep=crcr]{%
1	262.774393414388\\
10	262.774393414388\\
20	262.774393414388\\
30	262.774393414388\\
40	262.774393414388\\
50	262.774393414388\\
60	262.774393414388\\
70	262.774393414388\\
80  262.774393414388\\
90	262.774393414388\\
100	262.774393414388\\
};
\addlegendentry{Unclustered $b = 1$}

\legend{}
\end{axis}

\begin{axis}[%
    width=\fwidth,
    height=\fheight,
    at={(0\fwidth,0\fheight)},
    scale only axis,
    xmin=1,
    xmax=100,
    xtick={},
    ytick={},
    xticklabels={{}, {}, {},{}},
    yticklabels={},
    xtick style = {draw=none},
    ytick style = {draw=none},
    ymin=0.0001,
    ymax= 1,
 legend style={
        /tikz/every even column/.append style={column sep=0.2cm},
        at={(0.65, 0.02)},
        anchor=south, 
        draw=white!80!black, 
        font=\scriptsize,
        fill opacity=0.8
        },
    legend columns=2,
]
\addplot[color=mycolor2, only marks, very thick, mark size=3.0pt, mark=o, mark options={solid}] table[row sep=crcr]{%
1	-5 \\
};
\addlegendentry{CWC}
\addplot [color=black, solid, very thick]
  table[row sep=crcr]{%
1	-5\\
};
\addlegendentry{Unquantized}
\addplot[color=mycolor1, only marks, very thick, mark size=3.0pt, mark=x, mark options={solid}]table[row sep=crcr] {%
1	-5 \\
};
\addlegendentry{ICWC}
\addplot [color=black, dotted, very thick]
  table[row sep=crcr]{%
1	-5\\
};
\addlegendentry{1-bit quantized}
\addplot[color=mycolor7, only marks, very thick, mark size=3.0pt, mark=star, mark options={solid}] table[row sep=crcr] {%
1	-5 \\
};
\addlegendentry{Unclustered}
\addplot [color=black, dashed, very thick]
  table[row sep=crcr]{%
1	-5\\
};
\addlegendentry{2-bits quantized}

\end{axis}

\end{tikzpicture}%
    \caption{Average sum capacity as a function of the maximum number of clusters   $Z$, for $N_{\rm I}=3200$ and for different degrees of quantization of the phase shifts, and considering a \ac{plos} channel for the \ac{irs}-\acp{ue} links.}
    \label{fig:cap_vs_bits}
\end{figure}
\begin{figure}[t]
    \centering
    \setlength\fwidth{0.82\columnwidth}
    \setlength\fheight{0.5\columnwidth}
    \definecolor{mycolor2}{rgb}{1,0.2,0.2}%%ICWC
\definecolor{mycolor1}{rgb}{0,0,0.9}%%CWC
\definecolor{mycolor7}{rgb}{0.92900,0.69400,0.12500}
\pgfplotsset{every tick label/.append style={font=\scriptsize}}
\begin{tikzpicture}

\begin{axis}[%
     width=\fwidth,
    height=\fheight,
    at={(0\fwidth,0\fheight)},
    scale only axis,
    xlabel style={font=\footnotesize},
    ylabel style={font=\footnotesize},
    xmin=1,
    xmax=100,
    xtick={  1,  10,  20,  30,  40,  50,  60,  70,  80,  90, 100},
    xlabel={Maximum number of clusters $Z$},
    ymin=0.001,
    ymax= 0.42,
    %ymode=log,
    ylabel={$C_{95\%}$ [Mbit/s]},
    axis background/.style={fill=white},
    xmajorgrids,
    ymajorgrids,
    legend style={at={(0.65,0.02)}, anchor=south east, legend columns=2, legend cell align=left, align=left, font=\footnotesize, draw=white!15!black}
]

\addplot [color=mycolor1, very thick, mark size=3.0pt, mark=x, mark options={solid, mycolor1}]
  table[row sep=crcr]{%
1	0.00021353977713999\\
10	0.00885511846811477\\
20	0.013744379545304\\
30	0.0208065402208952\\
40	0.0277891151667394\\
50	0.0375180036653679\\
60	0.0733218322465392\\
70	0.21139661183481\\
80	0.339879349221883\\
90	0.394219418443468\\
100	0.39935086872942\\
};
\addlegendentry{ICWC}

\addplot [color=mycolor2, very thick, mark size=3.0pt, mark=o, mark options={solid, mycolor2}]
  table[row sep=crcr]{%
1	0.000201347131794128\\
10	0.00312580402520538\\
20	0.00573819025062837\\
30	0.00895436056303841\\
40	0.011225030668531\\
50	0.0131333117023435\\
60	0.0151559423415213\\
70	0.0175307211254688\\
80	0.0201655844663867\\
90	0.023766885578881\\
100	0.39935086872942\\
};
\addlegendentry{CWC}

\addplot [color=mycolor7, very thick, very thick, mark size=3.0pt, mark=star, mark options={solid}]
  table[row sep=crcr]{%
1	0.39935086872942\\
10	0.39935086872942\\
20	0.39935086872942\\
30	0.39935086872942\\
40	0.39935086872942\\
50	0.39935086872942\\
60	0.39935086872942\\
70	0.39935086872942\\
80	0.39935086872942\\
90	0.39935086872942\\
100	0.39935086872942\\
};
\addlegendentry{Unclustered}

\addplot [color=mycolor2, dotted, very thick, mark size=3.0pt, mark=o, mark options={solid}]
  table[row sep=crcr]{%
1	0.000466728077223696\\
10	0.00765883332664666\\
20	0.0125851396013992\\
30	0.0152030901129695\\
40	0.0170817992838275\\
50	0.0187564845839567\\
60	0.0202159260672763\\
70	0.0216377201854924\\
80	0.0234011592744046\\
90	0.0288696411004593\\
100	0.203415752070666\\
};
\addlegendentry{CWC $b = 1$}

\addplot [color=mycolor1, dotted,very thick, mark size=3.0pt, mark=x, mark options={solid}]
  table[row sep=crcr]{%
1	0.000643833612757151\\
10	0.00981382998033692\\
20	0.0147603892677734\\
30	0.016906865657916\\
40	0.0222814087667978\\
50	0.0372915896256938\\
60	0.0833263139498508\\
70	0.169697581311707\\
80	0.19604514626526\\
90	0.201878287905474\\
100	0.203415752070666\\
};
\addlegendentry{ICWC $b = 1$}

\addplot [color=mycolor7, dotted, very thick, mark size=3.0pt, mark=star, mark options={solid}]
  table[row sep=crcr]{%
1	0.203415752070666\\
10	0.203415752070666\\
20	0.203415752070666\\
30	0.203415752070666\\
40	0.203415752070666\\
50	0.203415752070666\\
60	0.203415752070666\\
70	0.203415752070666\\
80	0.203415752070666\\
90	0.203415752070666\\
100	0.203415752070666\\
};
\addlegendentry{Unclustered $b =1$}

\addplot [color=mycolor2, dashed, very thick, mark size=3.0pt, mark=x, mark options={solid}]
  table[row sep=crcr]{%
1	0.000320262409866478\\
10	0.00636537870333484\\
20	0.011880816946464\\
30	0.015283347237257\\
40	0.0185464478165411\\
50	0.0217703873883307\\
60	0.0242584925010448\\
70	0.0264371699351561\\
80	0.0301925562194067\\
90	0.0384861822242106\\
100	0.35792717771614\\
};
\addlegendentry{CWC $b =2$}

\addplot [color=mycolor1, dashed, very thick, mark size=3.0pt, mark=x, mark options={solid}]
  table[row sep=crcr]{%
1	0.000610619410365895\\
10	0.012580768894899\\
20	0.0193586213719456\\
30	0.0287425746457723\\
40	0.0371354085856087\\
50	0.0540549036529304\\
60	0.108074100336223\\
70	0.266125391637318\\
80	0.340646402144676\\
90	0.354921195225962\\
100	0.35792717771614\\
};
\addlegendentry{ICWC $b =2$}

\addplot [color=mycolor7, dashed, very thick, mark size=3.0pt, mark=star, mark options={solid}]
  table[row sep=crcr]{%
1	0.35792717771614\\
10	0.35792717771614\\
20	0.35792717771614\\
30	0.35792717771614\\
40	0.35792717771614\\
50	0.35792717771614\\
60	0.35792717771614\\
70	0.35792717771614\\
80	0.35792717771614\\
90	0.35792717771614\\
100	0.35792717771614\\
};
\addlegendentry{Unclustered $b =2$}

\legend{}
\end{axis}

\begin{axis}[%
    width=\fwidth,
    height=\fheight,
    at={(0\fwidth,0\fheight)},
    scale only axis,
    xmin=1,
    xmax=100,
    xtick={},
    ytick={},
    xticklabels={{}, {}, {},{}},
    yticklabels={},
    xtick style = {draw=none},
    ytick style = {draw=none},
    ymin=0.0001,
    ymax= 1,
 legend style={
        /tikz/every even column/.append style={column sep=0.2cm},
        at={(0.35, 0.52)},
        anchor=south, 
        draw=white!80!black, 
        font=\scriptsize,
        fill opacity=0.8
        },
    legend columns=2,
]
\addplot[color=mycolor2, only marks, very thick, mark size=3.0pt, mark=o, mark options={solid}] table[row sep=crcr]{%
1	-5 \\
};
\addlegendentry{CWC}
\addplot [color=black, solid, very thick]
  table[row sep=crcr]{%
1	-5\\
};
\addlegendentry{Unquantized}
\addplot[color=mycolor1, only marks, very thick, mark size=3.0pt, mark=x, mark options={solid}]table[row sep=crcr] {%
1	-5 \\
};
\addlegendentry{ICWC}
\addplot [color=black, dotted, very thick]
  table[row sep=crcr]{%
1	-5\\
};
\addlegendentry{1-bit quantized}
\addplot[color=mycolor7, only marks, very thick, mark size=3.0pt, mark=star, mark options={solid}] table[row sep=crcr] {%
1	-5 \\
};
\addlegendentry{Unclustered}
\addplot [color=black, dashed, very thick]
  table[row sep=crcr]{%
1	-5\\
};
\addlegendentry{2-bits quantized}

\end{axis}

\end{tikzpicture}%
    \caption{$95$\% percentile of the user capacity as a function of the maximum number of clusters   $Z$, for $N_{\rm I}=3200$ and for different degrees of quantization of the phase shifts, and considering a \ac{plos} channel for the \ac{irs}-\acp{ue} links.}
    \label{fig:quantile_vs_bits}
\end{figure}
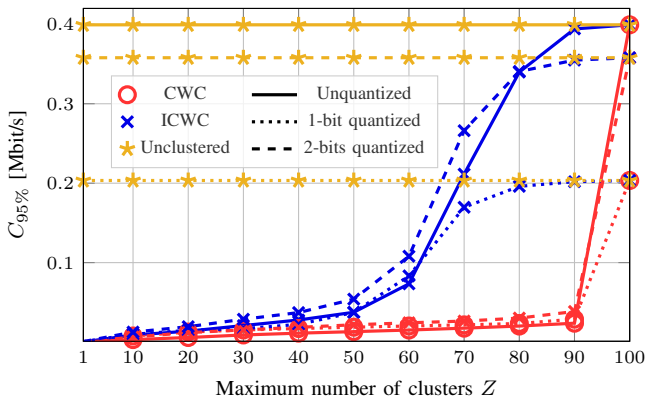

\vspace{2pt}\noindent\emph{Impact of quantization.} Figs.~\ref{fig:cap_vs_bits} and~\ref{fig:quantile_vs_bits} display the average sum capacity and the $95\%$ percentile, respectively, as a function of the maximum number of clusters $Z$ for \ac{cwc} and \ac{icwc}, and of the number of quantization bits $b$ of the phase shifts. 
Notice that energy and hardware constraints pose a limit to $b$~\cite{rivera2022optimization}, which implies restricting the infinite set of possible \ac{irs} configurations to a finite set of cardinality $2^{bN_{\rm I}}$. 
Moreover, the quantization constraint affects the beamforming capabilities of the \ac{irs} \cite{abeywickrama2020intelligent}, with negative implications for the resulting achievable sum capacity.
In \cite{rech2023downlink}, results were obtained considering that the quantization was performed only at the end of the clustering procedure. Here, instead, we assume that the quantization of the phase shifts is taken into account from the initial optimization stage.
The results reveal that the use of non-ideal phase shifters leads to a $30\%$ degradation in the sum capacity when using $b=1$ at the \ac{irs}, while the performance is close to the unquantized baseline if more quantization bits are used.
Furthermore, it is shown that the gap between quantized and the unquantized performance increases with $Z$. As a result, 1-bit quantization is sufficient to guarantee a performance comparable to the unquantized case with a small number of clusters, while more quantization bits are needed to achieve higher capacity.
In any case, we can conclude that our proposed capacity-based clustering algorithms are robust to phase-shift quantization. 

\begin{figure}[t]
    \centering
    \setlength\fwidth{0.82\columnwidth}
    \setlength\fheight{0.5\columnwidth}
    \definecolor{mycolor2}{rgb}{1,0.2,0.2}%%
\definecolor{mycolor1}{rgb}{0,0,0.9}
\definecolor{mycolor3}{rgb}{0.46600,0.67400,0.18800}%
\definecolor{mycolor7}{rgb}{0.92900,0.69400,0.12500}%
\definecolor{mycolor5}{rgb}{0.30100,0.74500,0.93300}%
\definecolor{mycolor6}{rgb}{0.49400,0.18400,0.55600}%
\definecolor{mycolor4}{rgb}{0.63500,0.07800,0.18400}%
\pgfplotsset{every tick label/.append style={font=\scriptsize}}
\begin{tikzpicture}
%%%%%%%%%%%%%%%%%%%%%%%%%%%%%%%%%%

\begin{axis}[%
    width=\fwidth,
    height=\fheight,
    at={(0\fwidth,0\fheight)},
    scale only axis,
    xlabel style={font=\footnotesize},
    ylabel style={font=\footnotesize},
    xmin=0.1,
    xmax=1,
    ymin=3500,
    ymax=55000,
    xtick={  0.1,  0.2,  0.3, 0.4, 0.5,  0.6,  0.7,  0.8,  0.9, 1},
    xlabel={$Z/K$},
    ylabel={Sum capacity $K\bar{C}$ [Mbit/s]},
    axis background/.style={fill=white},
    xmajorgrids,
    ymajorgrids,
        legend image post style={mark indices={}},
    legend style={
        /tikz/every even column/.append style={column sep=0.2cm},
        at={(0.5, 1.02)}, 
        anchor=south, 
        draw=white!80!black, 
        font=\footnotesize
        },
    legend columns=4,
]

% \addplot [color=mycolor5, very thick, dotted, mark size=3.0pt, mark=diamond*, mark options={solid, mycolor5}]
%   table[row sep=crcr]{%
% 0.1	7158.52457171352\\
% 0.2	9187.95734098891\\
% 0.3	10804.1875960404\\
% 0.4	12024.271793088\\
% 0.5	13084.5672883801\\
% 0.6	13858.325257917\\
% 0.7	14669.7117496\\
% 0.8	15460.2653956672\\
% 0.9	16141.2402404228\\
% 1	16782.4873258697\\
% };
% \addlegendentry{KM $K = 50$}

\addplot [color=mycolor6, very thick, dotted, mark size=3.0pt, mark=pentagon*, mark options={solid, mycolor6}]
  table[row sep=crcr]{%
0.1	4972.62550536484\\
0.2	7672.35058516874\\
0.3	9381.8170508526\\
0.4	10808.6538357438\\
0.5	12220.7622707602\\
0.6	13420.8310563712\\
0.7	14276.1701352295\\
0.8	15143.4573367123\\
0.9	15999.4932049858\\
1	16782.4873258697\\
};
\addlegendentry{HC $K = 50$}

\addplot [color=mycolor2, very thick, dotted, mark size=3.0pt, mark=o, mark options={solid, mycolor2}]
  table[row sep=crcr]{%
0.1	6259.2867659009\\
0.2	9037.2017613148\\
0.3	11256.1663762424\\
0.4	12910.8767632059\\
0.5	14145.0683221785\\
0.6	15056.3126665502\\
0.7	15718.9691054342\\
0.8	16191.923074003\\
0.9	16539.3942914684\\
1	16782.4873258697\\
};
\addlegendentry{CWC $K = 50$}

\addplot [color=mycolor1, very thick, dotted, mark size=3.0pt, mark=x, mark options={solid, mycolor1}]
  table[row sep=crcr]{%
0.1	3589.64383657103\\
0.2	5668.6541835302\\
0.3	7221.79507386142\\
0.4	8979.92041782268\\
0.5	10264.9479754787\\
0.6	11476.1756775995\\
0.7	12474.7878945357\\
0.8	13744.5149326159\\
0.9	15004.3644044722\\
1	16782.4873258697\\
};
\addlegendentry{ICWC $K = 50$}

% \addplot [color=mycolor3, very thick, dotted, mark size=3.0pt, mark=triangle*, mark options={solid, rotate=180, mycolor3}]table[row sep=crcr]{%
% 0.1	5420.13667963753\\
% 0.2	8646.62996145707\\
% 0.3	10960.7256035112\\
% 0.4	12729.2532623228\\
% 0.5	14023.2090414264\\
% 0.6	14974.2023292788\\
% 0.7	15686.1307542102\\
% 0.8	16170.4441113733\\
% 0.9	16531.5642675861\\
% 1	16782.4873258697\\
% };
% \addlegendentry{OSCBC $K = 50$}

% \addplot [color=mycolor5, very thick, mark size=3.0pt, mark=diamond*, mark options={solid, mycolor5}]
%   table[row sep=crcr]{%
% 0.10	7779.83732080269\\
% 0.20	12356.6685831694\\
% 0.30	15981.1932668768\\
% 0.40	19286.2674643542\\
% 0.50	22188.0711483525\\
% 0.60	25040.3247328401\\
% 0.70	27701.8934845287\\
% 0.80	30212.2691870657\\
% 0.90	32571.7076087751\\
% 1	34962.5593642553\\
% };
% \addlegendentry{KM $K = 100$}

\addplot [color=mycolor6, very thick, mark size=3.0pt, mark=pentagon*, mark options={solid, mycolor6}]
  table[row sep=crcr]{%
0.10	11292.5237968855\\
0.20	16683.0476313084\\
0.30	20567.2720543757\\
0.40	23677.8342633094\\
0.50	26430.8292456048\\
0.60	28584.0372512358\\
0.70	30443.0768920583\\
0.80	32054.3034112752\\
0.90	33524.8581783295\\
1	34962.5593642553\\
};
\addlegendentry{HC $K = 100$}

\addplot [color=mycolor2, very thick, mark size=3.0pt, mark=o, mark options={solid, mycolor2}]
  table[row sep=crcr]{%
0.10	14625.1344545728\\
0.20	20044.7073159671\\
0.30	24473.2697150523\\
0.40	27670.3626326244\\
0.50	30030.0582274394\\
0.60	31728.4325558418\\
0.70	32975.1795212581\\
0.80	33881.6310164981\\
0.90	34511.3220285055\\
1	34962.5593642553\\
};
\addlegendentry{CWC $K = 100$}

\addplot [color=mycolor1, very thick, mark size=3.0pt, mark=x, mark options={solid, mycolor1}]
  table[row sep=crcr]{%
0.10	9456.87304604159\\
0.20	13609.6494188705\\
0.30	17157.1973470307\\
0.40	20284.6458893204\\
0.50	22785.446794259\\
0.60	24801.792008807\\
0.70	26917.0203404248\\
0.80	29495.5409175835\\
0.90	32488.0336264252\\
1	34962.5593642553\\
};
\addlegendentry{ICWC $K = 100$}

% \addplot [color=mycolor3, very thick, mark size=3.0pt, mark=triangle*, mark options={solid, rotate=180, mycolor3}]
%   table[row sep=crcr]{%
% 0.10	12172.7908917765\\
% 0.20	18692.6040773053\\
% 0.30	23587.1625771207\\
% 0.40	27073.9181989123\\
% 0.50	29660.3964541797\\
% 0.60	31511.9655587025\\
% 0.70	32840.8235238309\\
% 0.80	33808.6670053488\\
% 0.90	34489.3216468099\\
% 1	34962.5593642553\\
% };
% \addlegendentry{OSCBC $K = 100$}

% \addplot [color=mycolor5, very thick, dotted, mark size=3.0pt, mark=diamond*, mark options={solid, mycolor5}]
%   table[row sep=crcr]{%
% 0.1	22910.1297159358\\
% 0.2	29715.5868285555\\
% 0.3	34780.1024846048\\
% 0.4	38021.1789388608\\
% 0.5	41162.6277576244\\
% 0.6	43349.0479357406\\
% 0.7	45869.2148433377\\
% 0.8	47731.3554605228\\
% 0.9	49583.1910268049\\
% 1	51333.382104275\\
% };
% \addlegendentry{KM $K = 150$}

\addplot [color=mycolor6, very thick, dashed, mark size=3.0pt, mark=pentagon*, mark options={solid, mycolor6}]
  table[row sep=crcr]{%
0.1	17092.1528605486\\
0.2	24623.4257328976\\
0.3	30265.6598232776\\
0.4	34786.2219244176\\
0.5	38771.1638935248\\
0.6	42116.8009455703\\
0.7	44601.4737948765\\
0.8	46914.0151602316\\
0.9	49099.3446937558\\
1	51333.382104275\\
};
\addlegendentry{HC $K = 150$}

\addplot [color=mycolor2, very thick, dashed, mark size=3.0pt, mark=o, mark options={solid, mycolor2}]
  table[row sep=crcr]{%
0.1	22613.1653281739\\
0.2	30636.0985237628\\
0.3	36446.3086549684\\
0.4	40838.1033077436\\
0.5	44275.4300047741\\
0.6	46664.385888016\\
0.7	48442.6851045046\\
0.8	49731.3411615368\\
0.9	50649.8455005196\\
1	51333.382104275\\
};
\addlegendentry{CWC $K = 150$}

\addplot [color=mycolor1, very thick, dashed, mark size=3.0pt, mark=x, mark options={solid, mycolor1}]
  table[row sep=crcr]{%
0.1	15066.7805883783\\
0.2	21046.6345381312\\
0.3	25930.0778832919\\
0.4	30691.0928797951\\
0.5	34711.3770357199\\
0.6	37817.2703433343\\
0.7	40815.9690027346\\
0.8	44569.8261707883\\
0.9	48190.7717442458\\
1	51333.382104275\\
};
\addlegendentry{ICWC $K = 150$}

% \addplot [color=mycolor3, very thick, dashed, mark size=3.0pt, mark=triangle*, mark options={solid, rotate=180, mycolor3}]table[row sep=crcr]{%
% 0.1	18373.5961553671\\
% 0.2	28061.2089805988\\
% 0.3	34917.9847806132\\
% 0.4	39831.9335508363\\
% 0.5	43654.9000187069\\
% 0.6	46250.8570781947\\
% 0.7	48187.2352016457\\
% 0.8	49565.535462766\\
% 0.9	50591.5768660543\\
% 1	51333.382104275\\
% };
% \addlegendentry{OSCBC $K = 150$}

\legend{}
\end{axis}

\begin{axis}[%
    width=\fwidth,
    height=\fheight,
    at={(0\fwidth,0\fheight)},
    scale only axis,
    xmin=1,
    xmax=100,
    xtick={},
    ytick={},
    xticklabels={{}, {}, {},{}},
    yticklabels={},
    xtick style = {draw=none},
    ytick style = {draw=none},
    ymin=0.0001,
    ymax= 1,
 legend style={
        /tikz/every even column/.append style={column sep=0.2cm},
        at={(0.26, 0.7)}, 
        anchor=south, 
        draw=white!80!black, 
        font=\scriptsize,
        fill opacity=0.8
        },
    legend columns=2,
]

% \addplot [color=mycolor5, only marks, very thick, mark size=3.0pt, mark=diamond*, mark options={solid, mycolor5}]
%   table[row sep=crcr]{%
% 1	-5\\
% };
% \addlegendentry{KM}

\addplot [color=mycolor6, only marks, very thick, mark size=3.0pt, mark=pentagon*, mark options={solid, mycolor6}] table[row sep=crcr] {%
1	-5 \\
};
\addlegendentry{HC}
\addplot [color=black, dotted, very thick]
  table[row sep=crcr]{%
1	-5\\
};
\addlegendentry{$K=50$}
\addplot[color=mycolor2, only marks, very thick, mark size=3.0pt, mark=o, mark options={solid}] table[row sep=crcr]{%
1	-5 \\
};
\addlegendentry{CWC}
\addplot [color=black, solid, very thick]
  table[row sep=crcr]{%
1	-5\\
};
\addlegendentry{$K=100$}
\addplot[color=mycolor1, only marks, very thick, mark size=3.0pt, mark=x, mark options={solid}] table[row sep=crcr] {%
1	-5 \\
};
\addlegendentry{ICWC}
\addplot [color=black, dashed, very thick]
  table[row sep=crcr]{%
1	-5\\
};
\addlegendentry{$K=150$}

% \addplot [color=mycolor3, only marks, very thick, mark size=3.0pt, mark=triangle*, mark options={solid, rotate=180, mycolor3}] table[row sep=crcr]{%
% 1	-5 \\
% };
% \addlegendentry{OSCBC}

\end{axis}

\end{tikzpicture}%  
    \caption{Sum capacity as a function of the maximum number of clusters over the number of \acp{ue} $Z/K$, for different values fo $K$, for an unquantized $40$H$\times80$V \ac{irs}, $K = \{50, 100, 150\}$, and considering a \ac{plos} channel for the \ac{irs}-\acp{ue} links. For readability, the results are shown without averaging over the \ac{tdma} frame length.} 
    \label{fig:sumcap_vs_ZK}
\end{figure}
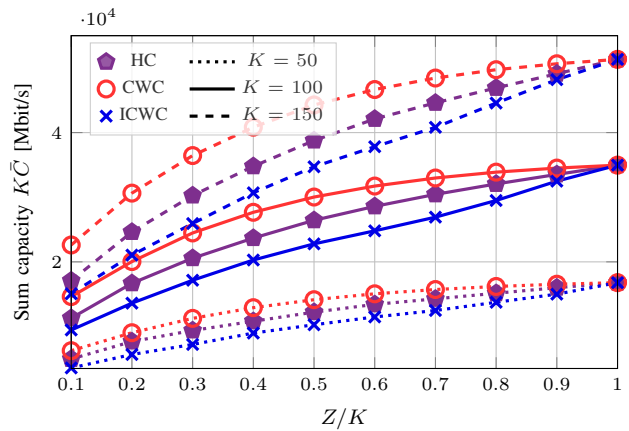

\begin{figure}[t]
    \centering
    \setlength\fwidth{0.82\columnwidth}
    \setlength\fheight{0.5\columnwidth}
    \input{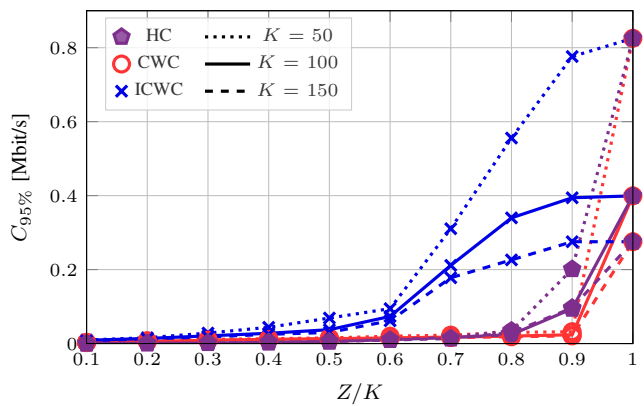}   
    \caption{$95$\% of the user capacity as a function of the maximum number of clusters over the number of \acp{ue} $Z/K$, for different values of $K$, for an unquantized $40$H$\times80$V \ac{irs}, $K = \{50, 100, 150\}$ \acp{ue}, and considering a \ac{plos} channel for the \ac{irs}-\acp{ue} links.}  
    \label{fig:quantcap_vs_ZK}
\end{figure}

\begin{figure}[t]
    \centering
    \setlength\fwidth{0.82\columnwidth}
    \setlength\fheight{0.5\columnwidth}
    \definecolor{mycolor2}{rgb}{1,0.2,0.2}%%
\definecolor{mycolor1}{rgb}{0,0,0.9}
\definecolor{mycolor3}{rgb}{0.46600,0.67400,0.18800}%
\definecolor{mycolor7}{rgb}{0.92900,0.69400,0.12500}%
\definecolor{mycolor5}{rgb}{0.30100,0.74500,0.93300}%
\definecolor{mycolor6}{rgb}{0.49400,0.18400,0.55600}%
\definecolor{mycolor4}{rgb}{0.63500,0.07800,0.18400}%
\pgfplotsset{every tick label/.append style={font=\scriptsize}}
\begin{tikzpicture}

%%%%%%%%%%%%%%%%%%%%%%%%%%%%%%%%%%

\begin{axis}[%
    width=\fwidth,
    height=\fheight,
    at={(0\fwidth,0\fheight)},
    scale only axis,
        xlabel style={font=\footnotesize},
    ylabel style={font=\footnotesize},
    xmin=20,
    xmax=160,
    xtick={20, 40, 60, 80, 100, 120, 140, 160, 180, 200},
    ymin=5,
    ymax=120,
    ytick={10, 20, 40, 60, 80, 100, 120, 140, 160},
    xlabel={Number of \acp{ue} $K$},
    ylabel={Min. number of IRS configurations $Z_{\rm min}$},
    axis background/.style={fill=white},
    xmajorgrids,
    ymajorgrids,
 legend style={
        /tikz/every even column/.append style={column sep=0.2cm},
        at={(0.15, 0.45)}, 
        anchor=south, 
        draw=white!80!black, 
        font=\scriptsize,
        fill opacity=0.8
        },
    legend columns=1,
]

\addplot [color=mycolor4, very thick, mark size=2.8pt, mark=square*, mark options={solid, mycolor4}]
  table[row sep=crcr]{%
20	13.8\\
40	28.4\\
60	41.7\\
80	53\\
100	65\\
120	78.3\\
140	86\\
160	98.4\\
180	106.2\\
200 115.7\\
};
\addlegendentry{KMed}

\addplot [color=mycolor5, very thick, mark size=3.0pt, mark=diamond*, mark options={solid, mycolor5}]
  table[row sep=crcr]{%
20	11.5\\
40	23.3\\
60	34.1\\
80	44.5\\
100	54.5\\
120	64.5\\
140	74\\
160	83.5\\
180	91.5\\
200 99.5\\
};
\addlegendentry{KM}

\addplot [color=mycolor6, very thick, mark size=3.0pt, mark=pentagon*, mark options={solid, mycolor6}]
  table[row sep=crcr]{%
20	13\\
40	25.4\\
60	37.4\\
80	49.4\\
100	62.2\\
120	73.2\\
140	83.7\\
160	93.6\\
180	104.1\\
200 115.0\\
};
\addlegendentry{HC}

\addplot [color=mycolor2, very thick, mark size=3.0pt, mark=o, mark options={solid, mycolor2}]
  table[row sep=crcr]{%
20	9.9\\
40	19.2\\
60	27.8\\
80	36.9\\
100	44\\
120	52.5\\
140	59.5\\
160	67\\
180	74.4\\
200 82.5\\
};
\addlegendentry{CWC}

\addplot [color=mycolor1, very thick, mark size=3.0pt, mark=x, mark options={solid, mycolor1}]
  table[row sep=crcr]{%
20	16.4\\
40	32.8\\
60	48.5\\
80	62.9\\
100	77.5\\
120	90.5\\
140	103.5\\
160	116\\
180	127\\
200 138\\
};
\addlegendentry{ICWC}

\addplot [color=mycolor3, very thick, mark size=3.0pt, mark=triangle*, mark options={solid, rotate=180, mycolor3}]
  table[row sep=crcr]{%
20	9.9\\
40	20\\
60	28.4\\
80	38.1\\
100	46.5\\
120	55\\
140	63\\
160	71\\
180	79\\
200 87\\
};
\addlegendentry{OSCBC}

\end{axis}
\end{tikzpicture}%   
    \caption{Minimum number of \ac{irs} configurations (clusters) $Z_{\rm min}$ to achieve 80\% of the maximum achievable sum capacity, for an unquantized $40$H$\times80$V \ac{irs}, and considering a \ac{plos} channel for the \ac{irs}-\acp{ue} links.}   
    \label{fig:minconfvsN}
\end{figure}
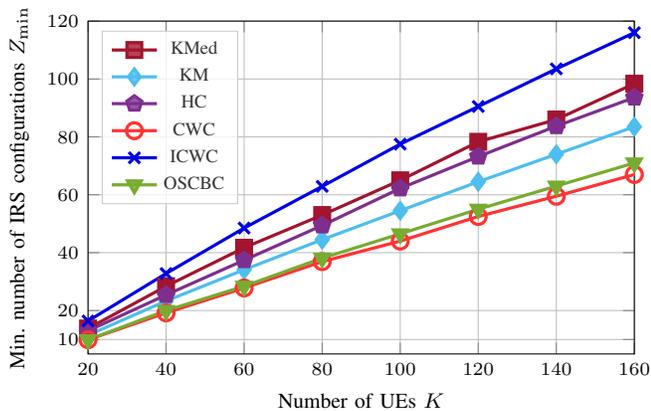

\vspace{2pt}\noindent\emph{Scalability.} Finally, we prove the scalability performance of the proposed clustering algorithms. 
To do so, we first show the performance of capacity-based clustering as a function of the number of \acp{ue} in Figs~\ref{fig:sumcap_vs_ZK} and~\ref{fig:quantcap_vs_ZK}. In particular, we compare \ac{cwc} and \ac{icwc} with \ac{hc} as a function of the ratio $K/N$, for an unquantized $40$H$\times80$V \ac{irs}, $K = \{50, 100, 150\}$ \acp{ue}, and considering a \ac{plos} channel for the \ac{irs}-\acp{ue} links.
The results are in line with the plots in Figs~\ref{fig:sumcap0bitlos} and \ref{fig:quantile0bitsplos}, which demonstrates the scalability of the proposed clustering techniques for different numbers of \acp{ue}.
Finally, Fig.~\ref{fig:minconfvsN} depicts the average minimum number of \ac{irs} configurations  $Z_{\rm min}$ needed to achieve 80\% of the maximum achievable sum capacity (``unclustered'' baseline) as a function of the number of \acp{ue} $K$ in the system.
Notably, we observe that \ac{cwc} and \ac{oscbc} are confirmed to be the best algorithms to optimize the sum capacity, even for a limited number of \ac{irs} configurations. For example, both solutions achieve 80\% of the maximum sum capacity with less than half the number of configurations than in the ``unclustered'' deployment. 
Moreover, we recognize the same trends as in the previous results. Specifically, capacity-based clustering outperforms distance-based clustering and requires fewer \ac{irs} reconfigurations to maximize the sum capacity (up to $-37\%$ considering \ac{cwc} vs. \ac{kmed}). Furthermore, the gap increases as the number of \acp{ue} increases.

\section{Conclusions}\label{sec:conclusions}
We considered a \ac{mimo} cellular network, in which a \ac{gnb} serving multiple \acp{ue} is assisted by an \ac{irs} acting as a relay. 
Notably, we considered practical constraints on the \ac{irs} reconfiguration period. We studied a \ac{tdma} scheduling for downlink transmissions, and formulated an optimization problem to maximize the average sum capacity, subject to a fixed number of \ac{irs} reconfigurations per time frame. We first discussed an iterative algorithm to obtain the optimal \ac{irs} configuration of each \ac{ue}. 
Then, we proposed clustering-based scheduling algorithms, which group \acp{ue} with similar (ideal) \ac{irs} configurations based on either a distance metric or the achievable capacity, to mitigate the performance degradation due to the constraint in the number of possible reconfigurations.
Different clustering algorithms were numerically evaluated in terms of computational complexity, sum capacity, and fairness under different channel conditions, as a function of the size of the \ac{irs} size and the number of users, and with or without quantization of phase shifts.
The results showed that capacity-based clustering outperforms distance-based clustering, and can achieve up to $85$\% of the sum capacity obtained in an ideal deployment (with no reconfiguration constraints), reducing by $50$\% the number of \ac{irs} reconfigurations.

\bibliographystyle{IEEEtran}
\bibliography{IEEEabrv,biblio_new}

\end{document}